\documentclass[showpacs,preprintnumbers,amsmath,amssymb,prb,twocolumn,floatfix]{revtex4}
\usepackage[dvips]{graphics,color}
\usepackage{dcolumn}
\begin{document}
\title{Non-Adiabatic Spin Transfer Torque in Real Materials}
\author{Ion Garate$^1$, K. Gilmore$^{2,3}$, M. D. Stiles$^2$, and A.H. MacDonald$^1$}
\affiliation{$^1$Department of Physics, The University of Texas at Austin, Austin, TX 78712}
\affiliation{$^2$Center for Nanoscale Science and Technology, National
  Institute of Standards and Technology, Gaithersburg, MD 20899-8412 }
\affiliation{$^3$Maryland NanoCenter, University of Maryland, College Park, MD, 20742}
\date{\today}
\begin{abstract}
The motion of simple domain walls and of more complex magnetic textures in the presence of a 
transport current is described by the Landau-Lifshitz-Slonczewski (LLS) equations.  Predictions of the 
LLS equations depend sensitively on the ratio between the dimensionless material parameter $\beta$ which 
characterizes non-adiabatic spin-transfer torques and the Gilbert damping parameter $\alpha$. 
This ratio has been variously estimated to be close to 0, close to 1, 
and large compared to 1.  By identifying $\beta$ as the influence of a transport current on $\alpha$, we derive a 
concise, explicit and relatively simple expression which relates $\beta$ to the band structure and Bloch state lifetimes of a magnetic metal. Using this expression we demonstrate that intrinsic spin-orbit interactions lead to 
intra-band contributions to $\beta$ which are often dominant and can be (i) estimated with some confidence and  
(ii) interpreted using the ``breathing Fermi surface'' model. 
\end{abstract}
\maketitle

\section{Introduction}
An electric current can
influence the magnetic state of a ferromagnet by  exerting a
\emph{spin transfer torque}(STT) on the magnetization.\cite{berger_old,berger,slonc} 
This effect occurs whenever currents travel through non-collinear
magnetic systems and is therefore promising for magnetoelectronic applications.
Indeed, STT's have already been exploited in
a number of technological devices.\cite{applications} Partly for
this reason and partly because the quantitative description of order
parameter manipulation by out-of-equilibrium quasiparticles poses
great theoretical challenges, the study of the STT effect  has
developed into a major research subfield of spintronics. 

Spin transfer torques are important in both magnetic multilayers,
where the magnetization changes abruptly,\cite{reviews_mm} and in
magnetic nanowires, where the magnetization changes
smoothly.\cite{reviews} Theories of the STT in systems with smooth magnetic textures
identify two different types of spin transfer.
On one hand, the adiabatic or Slonczewski\cite{slonc} torque results
when quasiparticle spins follow the underlying magnetic landscape
adiabatically. It can be mathematically expressed as $(\textbf{v}_{\rm
s}\cdot\nabla){\bf s_0}$, where ${\bf s_0}$ stands for the
magnetization and $\textbf{v}_{\rm s}$ is the ``spin velocity'', which is 
proportional to the charge drift velocity, and hence to the current and the applied electric field. 
The microscopic physics of the Slonczewski spin-torque is thought to be 
well understood\cite{reviews,reviews_mm,stiles}, at least\cite{alvaro} in systems with weak spin-orbit coupling.
A simple angular momentum conservation argument argues that in the absence of 
spin-orbit coupling ${\bf v_{\rm s}}=\sigma_s {\bf E} /e s_0$, where $s_0$ is the magnetization, $\sigma_s$ is the spin conductivity and ${\bf E}$ is the electric field. 
However, spin-orbit coupling plays an essential role in real magnetic materials and hence the validity of this simple expression for $v_s$ needs to be tested by more rigorous calculations.  

The second spin transfer torque in continuous media, $\beta{\bf s_0}\times(\textbf{v}_{\rm s}
\cdot\nabla){\bf s_0}$,
acts in the perpendicular direction and is frequently referred to as
the non-adiabatic torque.\cite{nonadiabatic} Unfortunately, the name is a misnomer in the present context.  There are two contributions that have the preceeding form.  The first is truly non-adiabatic and occurs in systems in which the magnetization varies too rapidly in space for the spins of the transport electrons to follow the local magnetization direction as they traverse the magnetization texture.  For wide domain walls, these effects are expected to be small.\cite{xiao}  The contribution of interest in this paper is a dissipative contribution that occurs in the adiabatic limit.  The adiabatic torque discussed above is the reactive contribution in this limit.  As we discuss below, processes that contribute to magnetic damping, whether they derive from spin-orbit coupling or spin-dependent scattering, also give a spin-transfer torque parameterized by $\beta$ as above.  In this paper, we follow the common convention and refer to this torque as non-adiabatic.  However, it should be understood that it is a dissipative spin transfer torque that is present in the adiabatic limit.

The non-adiabatic torque
plays a key role in current-driven domain wall dynamics, where the
ratio between $\beta$ and the Gilbert parameter $\alpha$ can determine
the velocity of domain walls under the influence of a transport
current. 
There is no consensus on its magnitude of the parameter $\beta$.\cite{domain, reviews} Although there have a few theoretical studies\cite{tserkovnyak,tatara,rembert} of the STT in toy models, the relationship between
toy model STT's and STT's in either transition metal ferromagnets or
ferromagnetic semiconductors is far from clear.
As we will discuss the toy models most often studied neglect spin-orbit interactions 
in the band-structure of the perfect crystal, {\em intrinsic} spin-orbit interactions, which can alter STT physics qualitatively.

The main objectives of this paper are (i) to shed new light on the physical meaning of the non-adiabatic STT by relating it to the change in magnetization damping due to a transport current, (ii) to derive a concise formula that can be used to evaluate $\beta$ in real materials from first principles and (iii) to demonstrate
that  $\alpha$ and $\beta$ have the same qualitative dependence on disorder (or temperature), even though their ratio depends on the details of the band structure. 
As a byproduct of our theoretical study, we find that the expression for $v_s$ in terms of the spin conductivity may not always be accurate in materials with strong spin-orbit coupling.

We begin in Section II by reviewing and expanding on microscopic 
theories of $\alpha$, $\beta$ and $v_s$. 
In short, our microscopic approach quantifies how the micromagnetic energy of an inhomogeneous ferromagnet is altered in response to 
external rf fields and dc transport currents which drive the magnetization direction away from local equilibrium.  These effects are captured by the spin transfer torques, damping torques, and 
effective magnetic fields that appear in the LLS equation.
By relating magnetization dynamics to effective magnetic fields, we derive explicit expressions for  $\alpha$,$\beta$ and $v_s$ in terms of microscopic parameters.
Important contributions to these materials parameters can be understood in clear physical terms using the breathing Fermi surface model.\cite{breathing} 
Readers mainly interested in a qualitative explanation for our findings may skip directly to Section VIII where 
we discuss of our main results in that framework. 
Regardless of the approach, the non-adiabatic STT can be understood as the change in the Gilbert damping contribution
to magnetization dynamics when the Fermi sea quasiparticle distribution function is altered by the transport electric field. 
The outcome of this insight is a concise analytical formula for $\beta$ which is simple enough that it can be conveniently 
combined with first-principles electronic structure calculations to predict $\beta$-values in particular materials.\cite{betaII}

In Sections III, IV and V we apply our expression for $\beta$ to model ferromagnets. In Section III we perform a necessary reality check
by applying our theory of $\beta$ to the parabolic band Stoner ferromagnet, the only model for which more rigorous 
fully microscopic calculations\cite{tatara,rembert} of $\beta$ have been completed. 
Section IV is devoted to the study of a two-dimensional electron-gas ferromagnet with Rashba spin-orbit interactions.
Studies of this model provide a qualitative indication of the influence of intrinsic spin-orbit interactions on the non-adiabatic STT. 
We find that, as in the microscopic theory\cite{alphaI,alphaII} for $\alpha$, spin-orbit interactions induce intra-band contributions to $\beta$ which are proportional to the quasiparticle lifetimes. These considerations carry over to the more sophisticated 4-band spherical model that we analyze in Section V; there our calculation is tailored to (Ga,Mn)As. We show that intra-band (conductivity-like) contributions are prominent in the 4-band model for experimentally relevant scattering rates. 

Section VI discusses the phenomenologically important $\alpha/\beta$ ratio for real materials. Using our analytical results derived in Section II (or Section VIII) we are able to reproduce and extrapolate trends expected from toy models which indicate that $\alpha/\beta$ should vary across materials in 
approximately the same way as the ratio between the itinerant spin density and the total spin density.  We also 
suggest that $\alpha$ and $\beta$ may have the opposite signs in systems with both hole-like and electron-like carriers.
We present concrete results for (Ga,Mn)As, where we obtain $\alpha/\beta\simeq 0.1$. 
This is reasonable in view of the weak spin polarization and the strong spin-orbit coupling of valence band holes in this material.     

Section VII describes the generalization of the \emph{torque-correlation} formula employed in \emph{ab-initio} calculations of the Gilbert damping to the case of the non-adiabatic spin-transfer torque. The torque correlation formula incorporates scattering of quasiparticles simply by introducing a phenomenological lifetime for the Bloch states and 
assumes that the most important electronic transitions occur between states near the Fermi surface in the same band.  
Our ability to make quantitative predictions based on this formula is limited mainly 
by an incomplete understanding of Bloch state scattering processes in real ferromagnetic materials.  
These simplifications give rise to ambiguities and inaccuracies that we dissect in Section VII.
Our assessment indicates that the torque correlation formula for $\beta$ is most accurate at low disorder and relatively weak spin-orbit interactions.

Section VIII restates and complements the effective field calculation explained in Section II. 
Within the adiabatic approximation, the instantaneous energy of a ferromagnet may be written in terms of the instantaneous occupation factors of quasiparticle states. We determine the effect of the external perturbations on the occupation factors by combining the relaxation time approximation and the master equation.  In this way we recover the results of Section II 
and are able to interpret the intra-band contributions to $\beta$ in terms of a generalized breathing Fermi surface picture.

Section IX contains a brief summary which concludes this work. 

\section{Microscopic Theory of $\alpha$, $\beta$ and $v_s$}
The Gilbert damping parameter $\alpha$, the non-adiabatic spin transfer torque coefficient $\beta$ and the ``spin velocity'' ${\bf v_s}$ appear in the generalized Landau-Lifshitz-Gilbert expression for collective magnetization dynamics of a ferromagnet under the influence of an electric current:
\begin{equation}
\label{eq:dynamics}
\left(\partial_{t}+\textbf{v}_{\rm s}\cdot\nabla\right)\hat{\Omega}-\hat{\Omega}\times {\cal H}_{\rm eff}=-\alpha\hat{\Omega}\times\partial_{t}\hat{\Omega}-\beta\hat{\Omega}\times (\textbf{v}_{\rm s}\cdot\nabla)\hat{\Omega}.
\end{equation}
In Eq.~(\ref{eq:dynamics}) ${\cal H}_{\rm eff}$ is an effective
magnetic field which we elaborate on below 
and $\hat{\Omega}={\bf s_0}/s_0\simeq(\Omega_{x},\Omega_{y},1-(\Omega_x^2+\Omega_y^2)/2)$ is the direction of the magnetization.\cite{convention}  Eq.~(\ref{eq:dynamics}) describes the slow dynamics of smooth magnetization textures in 
the presence of a weak electric field which induces transport currents.
It explicitly neglects the dynamics of the magnetization magnitude which is implicitly assumed to be 
negligible.  For small deviations from the easy direction (which we take to be the $\hat{z}$-direction), it reads
\begin{eqnarray}
\label{eq:h_eff}
{\cal H}_{{\rm eff},x}&=&\left(\partial_t+{\bf v}_{\rm s}\cdot\nabla\right)\Omega_y+\left(\alpha\partial_t+\beta{\bf v}_{\rm s}\cdot\nabla\right)\Omega_x\nonumber\\
{\cal H}_{{\rm eff},y}&=&-\left(\partial_t+{\bf v}_{\rm s}\cdot\nabla\right)\Omega_x+\left(\alpha\partial_t+\beta{\bf v}_{\rm s}\cdot\nabla\right)\Omega_y
\end{eqnarray} 
The gyromagnetic ratio has been absorbed into the units of the field ${\cal H}_{\rm eff}$ so that this quantity has inverse time units.  We set $\hbar=1$ throughout. 

In this section we relate the $\alpha$, $\beta$ and ${\bf v_s}$ parameters 
to microscopic features of the ferromagnet by considering the transverse total spin response function.
For a technically more accessible (yet less rigorous) theory of $\alpha$ and $\beta$ we refer to Section VIII. The transverse spin response function studied here describes the change in the micromagnetic energy due to the departure of the magnetization away from its equilibrium direction, where equilibrium is characterized by the absence of currents and external rf fields. This change in energy defines an effective magnetic field which may then be identified with Eq.~(\ref{eq:h_eff}), thereby allowing us to microscopically determine $\alpha$,$\beta$ and $v_s$.
To first order in frequency $\omega$, wave vector ${\bf q}$ and electric field, the transverse spin response function is given by 
\begin{widetext}
\begin{equation}
\label{eq:linear response}
S_0\hat{\Omega}_{a} =\sum_{b}\chi_{a,b}{\cal H}_{{\rm ext},b}
\simeq\sum_b\left[\chi_{a,b}^{(0)}+\omega\chi_{a,b}^{(1)}+({\bf
    v}_{\rm s}\cdot{\bf q})\chi_{a,b}^{(2)}\right]{\cal H}_{{\rm ext},b}
\end{equation}
where $a,b\in\{x,y\}$, ${\cal H}_{\rm ext}$ is the external magnetic field with frequency $\omega$ and wave vector ${\bf q}$, $S_0=s_0 V$ is the total spin of the ferromagnet ($V$ is the sample volume), and $\chi$ is the transverse spin-spin response function in the presence of a uniform time-independent electric field:
\begin{equation}
\label{eq:chi}
\chi_{a,b}({\bf q},\omega; {\bf v}_{\rm s})=i\int_{0}^{\infty} dt \int d{\bf r} \exp(i\omega t-i{\bf q}\cdot{\bf r})\langle\left[S^a({\bf r},t),S^b({\bf 0},0)\right]\rangle.
\end{equation}
\end{widetext}
In Eq.~(\ref{eq:linear response}), $\chi^{(0)}=\chi({\bf q=0},\omega=0; {\bf E=0})$ describes the spin response to a constant, uniform external magnetic field in absence of a current, $\chi^{(1)}=\lim_{\omega\to 0}\chi({\bf q=0},\omega; {\bf E=0})/\omega$ characterizes the spin response to a time-dependent, uniform external magnetic field in absence of a current, and $\chi^{(2)}=\lim_{q,v_s\to 0}\chi({\bf q},\omega=0; {\bf E})/{\bf q}\cdot{\bf v}_{\rm s}$ represents the spin response to a constant, non-uniform external magnetic field combined with a constant, uniform electric field {\bf E} . 
Note that first order terms in {\bf q} are allowed by symmetry in presence of an electric field. 
In addition, $\langle \rangle$ is a thermal and quantum mechanical average over states that describe a uniformly magnetized, current carrying ferromagnet. 

The approach underlying Eq.~(\ref{eq:linear response}) comprises a linear response theory with respect to an inhomogeneous magnetic field {\em followed by} a linear response theory with respect to an electric field. 
Alternatively, one may treat the electric and magnetic perturbations on an equal footing without predetermined ordering; for further considerations on this matter we refer to Appendix A.  

In the following we emulate and appropriately generalize a procedure outlined elsewhere.\cite{alphaI} 
First, we recognize that in the static limit and in absence of a current the transverse magnetization responds to the external magnetic field by adjusting its orientation to minimize the total energy including the internal energy $E_{int}$ and the energy due to coupling with the external magnetic field, $E_{ext}=-S_0\hat\Omega\cdot{\cal H}_{\rm ext}$. 
It follows that $\chi^{(0)}_{a,b}=S_0^2 [\partial^2 E_{\rm int}/\partial\hat\Omega_a\partial\hat\Omega_b]^{-1}$ and thus ${\cal H}_{\rm int, a}=-(1/S_0)\partial E_{\rm int}/\partial\hat\Omega_a=-S_0 [\chi^{(0)}]_{a,b}^{-1}\hat\Omega_b$, where ${\cal H}_{\rm int}$ is the internal energy contribution to the effective magnetic field. Multiplying  Eq.~(\ref{eq:linear response}) on the left by $[\chi^{(0)}]^{-1}$ and using ${\cal H}_{\rm eff}={\cal H}_{\rm int}+{\cal H}_{\rm ext}$ we obtain a formal equation for ${\cal H}_{\rm eff}$: 
\begin{equation}
\label{eq:h_eff_2}
{\cal H}_{{\rm eff},a}=\sum_b \left[{\cal L}_{a,b}^{(1)} \partial_t + {\cal L}_{a,b}^{(2)} ({\bf v}_{\rm s}\cdot\nabla)\right]\hat{\Omega}_b,
\end{equation}
where
\begin{eqnarray}
\label{eq:L}
{\cal L}^{(1)}&=&-i S_0[\chi^{(0)}]^{-1}\chi^{(1)}[\chi^{(0)}]^{-1}\nonumber\\
{\cal L}^{(2)}&=&i S_0[\chi^{(0)}]^{-1}\chi^{(2)}[\chi^{(0)}]^{-1}.
\end{eqnarray}
Identifying of Eqs.~(\ref{eq:h_eff_2}) and ~(\ref{eq:h_eff}) results in concise microscopic expressions for $\alpha$ and $\beta$ and ${\bf v}_{\rm s}$:
\begin{eqnarray}
\label{eq:alpha_beta def}
\alpha&=&{\cal L}_{x,x}^{(1)}={\cal L}^{(1)}_{y,y}\nonumber\\
\beta&=&{\cal L}_{x,x}^{(2)}={\cal L}^{(2)}_{y,y}\nonumber\\
1&=& {\cal L}^{(2)}_{x,y}\implies {\bf v}_{\rm s}\cdot{\bf q} =i S_0\left[(\chi^{(0)})^{-1}\chi(\chi^{(0)})^{-1}\right]_{x,y}.
\end{eqnarray}
In the third line of Eq.~(\ref{eq:alpha_beta def}) we have combined the second line of Eq.~(\ref{eq:L}) with $\chi^{(2)}=\chi/({\bf v}_{\rm s}\cdot {\bf q})$. 

When applying Eq.~(\ref{eq:alpha_beta def}) to realistic conducting ferromagnets, one must invariably adopt a self-consistent mean-field (Stoner) theory description of the magnetic state derived within a spin-density-functional theory (SDFT) framework.\cite{vignalesdftdynamics,sdft} In SDFT the transverse spin response function is expressed in terms of Kohn-Sham quasiparticle response to both external and induced magnetic fields; this allows us to transform\cite{alphaI} Eq.~(\ref{eq:alpha_beta def}) into 
\begin{eqnarray}
\label{eq:alpha_beta_SDFT}
\alpha&=&\frac{1}{S_0}\lim_{\omega\to 0}\frac{\text{Im}[\tilde{\chi}_{+,-}^{\rm QP}({\bf q}=0,\omega,{\bf E}=0)]}{\omega}\nonumber\\
\beta&=&-\frac{1}{S_0}\lim_{{\bf v}_{\rm s},{\bf q}\to 0}\frac{\text{Im}[\tilde{\chi}_{+,-}^{\rm QP}({\bf q},\omega=0,{\bf E})]}{{\bf q}\cdot{\bf v}_{\rm s}}\nonumber\\
{\bf v}_{\rm s}\cdot{\bf q} &=&-\frac{1}{S_0}\text{Re}[\tilde{\chi}_{+,-}^{\rm QP}({\bf q},\omega=0,{\bf E})],
\end{eqnarray}
where we have used\cite{caveat_0} $\chi^{(0)}_{a,b}=\delta_{a,b} S_0/\bar\Delta$ and
\begin{eqnarray}
\label{eq:chi_QP}
\tilde{\chi}^{\rm QP}_{+,-}({\bf q},\omega;{\bf
  E})&=&\frac{1}{2}\sum_{i,j}\frac{f_{j}-f_{i}}{\epsilon_{i}-\epsilon_{j}-\omega-i\eta}
\nonumber\\
&&\langle j|S^+\Delta_0({\bf r}) e^{i{\bf q}\cdot{\bf r}}|i\rangle \langle i|S^-\Delta_0({\bf r})e^{-i{\bf q}\cdot{\bf r}}|j\rangle \nonumber\\
\end{eqnarray}
is the quasiparticle response to changes in the  
direction of the exchange-correlation effective magnetic field.\cite{xx_vs_pm} 
 To estimate 
$\beta$ this response function should be evaluated in the presence of an electric current. 
In the derivation of Eq.~(\ref{eq:alpha_beta_SDFT}) we have made use of the fact that $\chi^{(1)}_{x,x}$ and $\chi^{(2)}_{x,x}$ are purely imaginary, whereas $\chi^{(2)}_{x,y}$ is purely real; this can be verified mathematically through $S^{\pm}=S_x\pm i S_y$. Physically, ``Im'' and ``Re'' indicate that the Gilbert damping and the non-adiabatic STT are dissipative while the adiabatic STT is reactive. 
Furthermore, in the third line it is implicit that we expand $\text{Re}[\tilde{\chi}^{\rm QP}]$ to first order in $q$ and $E$. 

In Eq.~(\ref{eq:chi_QP}),  $S^{\pm}$ is the spin-rising/lowering operator, $|i\rangle$, $\epsilon_{i}$  and $f_{i}$ are the Kohn-Sham eigenstates, eigenenergies and Fermi factors in presence of spin-dependent disorder, and 
$\Delta_0({\bf r})$ is the difference in the magnetic ground state between the majority spin and minority spin exchange-correlation potential - the spin-splitting potential.  This quantity is always 
spatially inhomogeneous at the atomic scale and is typically larger in atomic regions than in 
interstitial regions.  Although the spatial dependence of $\Delta_0({\bf r})$ plays a crucial role 
in realistic ferromagnets, we replace it by a phenomenological {\em constant} $\Delta_0$ in the toy models 
we discuss below.

Our expression of $v_s$ in terms of the transverse spin response function may be unfamiliar to readers familiar with the argument given in the introduction of this paper in which $v_s$ is determined by the divergence in spin current.  This argument is based on the assumption that the (transverse) angular momentum lost by spin polarized electrons traversing an inhomogeneous ferromagnet is transferred to the magnetization. However, this assumption fails when spin angular momentum is not conserved as it is not in the presence of spin-orbit coupling.  In general, part of the transverse spin polarization lost by the current carrying quasiparticles
is transferred to the lattice rather than to collective magnetic degrees of freedom\cite{alvaro} when spin-orbit interactions are present. It is often stated that the physics of spin non-conservation is captured by the non-adiabatic STT; however, the non-adiabatic STT {\em per se} is limited to dissipative processes and cannot describe the changes in the reactive spin torque due to spin-flip events. Our expression in terms of the transverse spin response function does not rely on spin conservation, and while it agrees with the conventional picture\cite{joaquin} in simplest cases (see below), it departs from it when {\em e.g.} intrinsic spin-orbit interactions are strong.       

In this paper we incorporate the influence of an electric field by simply shifting the Kohn-Sham orbital occupation factors 
to account for the energy deviation of the distribution function in a drifting Fermi sea: 
\begin{equation}
f_i\simeq f^{(0)}(\epsilon_i +V_i)\simeq f^{(0)}(\epsilon_i)+V_i \partial f^{(0)}/\partial \epsilon_{i}
\end{equation}
where $V_i$ is the effective energy shift for the $i$-th eigenenergy due to acceleration
between scattering events by an electric field and $f^{(0)}$ is the equilibrium Fermi factor.
This approximation to the steady-state induced by an external electric field is 
known to be reasonably accurate in many circumstances, for example in theories of 
electrical transport properties, and it can be used\cite{joaquin} to provide a microscopic 
derivation of the adiabatic spin-transfer torque.  As we discuss below, this 
{\em ansatz} provides a result for $\beta$ which is sufficiently simple that it can be combined  
with realistic {\em ab initio} electronic structure calculations to estimate $\beta$ values in 
particular magnetic metals.  We support this {\em ansatz} by demonstrating that 
it agrees with full non-linear response calculations in the case of toy models 
for which results are available.  

Using the Cauchy identity, $1/(x-i\eta)=1/x+i\pi\delta(x)$, and $\partial f^{(0)}/\partial\epsilon\simeq -\delta(\epsilon)$ we obtain
\begin{widetext}
\begin{eqnarray}
\label{eq:Im_chi}
\text{Im}[\tilde{\chi}_{+,-}^{\rm QP}]&\simeq&\frac{\pi}{2}\sum_{i,j}\left[\omega-V_{j,i}\right]|\langle j|S^+ \Delta_{0}({\bf r})\, e^{i {\bf q}\cdot {\bf r}}|i\rangle|^2\delta(\epsilon_{i}-\epsilon_{\rm F})\; \delta(\epsilon_{j}-\epsilon_{\rm F})\nonumber\\
\text{Re}[\tilde{\chi}_{+,-}^{\rm QP}]&\simeq& -\frac{1}{2}\sum_{i,j}|\langle j|S^+ \Delta_0({\bf r}) e^{i{\bf q}\cdot{\bf r}}|i\rangle|^2  \; \frac{V_j\delta(\epsilon_j-\epsilon_{\rm F})-V_i\delta(\epsilon_i-\epsilon_{\rm F})}{\epsilon_i-\epsilon_j}
\end{eqnarray}
\end{widetext}
where we have defined the difference in transport deviation energies by 
\begin{equation}
\label{eq:V_nn}
V_{j,i}\equiv V_j-V_i.
\end{equation}
In the first line of Eq.~(\ref{eq:Im_chi}), the two terms within the
square brackets correspond to the energy of particle-hole excitations
induced by radio frequency magnetic and static electric fields, respectively. The imaginary part selects scattering processes that relax the spin of the particle-hole pairs mediated 
either by phonons or by magnetic impurities.\cite{heinrich} Substituting Eq.~(\ref{eq:Im_chi}) into Eq.~(\ref{eq:alpha_beta_SDFT}) we can readily extract $\alpha$, $\beta$ and ${\bf v}_{\rm s}$:
\begin{widetext}
\begin{eqnarray}
\label{eq:alpha_beta_SDFT_2}
\alpha&=&\frac{\pi}{2 S_0}\sum_{i,j}|\langle j|S^+ \Delta_0({\bf r})|i\rangle|^2\delta(\epsilon_{i}-\epsilon_{\rm F})\delta(\epsilon_{j}-\epsilon_{\rm F})\nonumber\\
\beta&=&\lim_{q,v_s\to 0}\frac{\pi}{2 S_0 {\bf q}\cdot{\bf v}_{\rm s}}\sum_{i,j} |\langle j|S^+ \, \Delta_{0}({\bf r}) \, e^{i{\bf q}\cdot{\bf r}}|i\rangle|^2 \; V_{j,i} \; \delta(\epsilon_{i}-\epsilon_{\rm F})\delta(\epsilon_{j}-\epsilon_{\rm F})\nonumber\\
{\bf v}_{\rm s}\cdot{\bf q}&=&\frac{1}{2 S_0}\sum_{i,j}|\langle j|S^+ \Delta_0({\bf r}) e^{i{\bf q}\cdot{\bf r}}|i\rangle|^2 \frac{V_j\delta(\epsilon_j-\epsilon_{\rm F})-V_i\delta(\epsilon_i-\epsilon_{\rm F})}{\epsilon_i-\epsilon_j}
\end{eqnarray}
\end{widetext}
where we have assumed a uniform precession mode for the Gilbert damping. 

Eq.~(\ref{eq:alpha_beta_SDFT_2}) and Eq.~(\ref{eq:Im_chi}) identify the non-adiabatic STT  as a \emph{correction} to the Gilbert damping in 
the presence of an electric current; in other words, the magnetization damping at finite current is given by the sum of the Gilbert damping and the non-adiabatic STT. We feel that this simple interpretation of the non-adiabatic spin-transfer torque has not received sufficient emphasis in the literature.

Strictly speaking the influence of a transport current on magnetization dynamics should be calculated by considering non-linear response of 
transverse spin to both effective magnetic fields and the external electric field which drives the transport current.  
Our approach, in which we simply alter the occupation probabilities which appear in the transverse spin response function 
is admittedly somewhat heuristic.  We demonstrate below that it gives approximately the same result as the complete 
calculation for the case of the very simplistic model for which that complete calculation has been carried out.
 
In Eq.~(\ref{eq:alpha_beta_SDFT_2}), the eigenstates indexed by $i$ are not Bloch states of a periodic potential but instead the eigenstates of the Hamiltonian that includes all of the static disorder.  Although Eq.~(\ref{eq:alpha_beta_SDFT_2}) provides  
compact expressions valid for arbitrary metallic ferromagnets, its 
practicality is hampered by the fact that the characterization of disorder is normally not precise enough to permit a reliable solution of the Kohn-Sham equations with arbitrary impurities. An approximate yet more tractable treatment of disorder consists of the following steps: (i) replace the actual eigenstates of the disordered system by Bloch eigenstates corresponding to a pure crystal, {\em e.g.} $|i\rangle\to|{\bf k},a\rangle$, where ${\bf k}$ is the crystal momentum and $a$ is the band index of the perfect crystal; (ii) switch $V_i$ to $V_a=\tau_{{\bf k},a}{\bf v}_{{\bf k},a}\cdot e{\bf E}$, where $\tau$ is the Bloch state lifetime and ${\bf v}_{{\bf k},a}=\partial \epsilon_{{\bf k} ,a}/\partial {\bf k}$ is the quasiparticle group velocity, (iii) substitute the $\delta(\epsilon_{{\bf k},a}-\epsilon_{\rm F})$ spectral function of a Bloch state by a broadened spectral function evaluated at the Fermi energy: $\delta(\epsilon_{{\bf k},a}-\epsilon_{\rm F})\to A_{a}({\epsilon_{\rm F},\bf k})/(2\pi)$, where 
\begin{equation}
\label{eq:lorentzian}
A_{a}(\epsilon_{\rm F},{\bf k})=\frac{\Gamma_{\textbf{k},a}}{(\epsilon_{\rm F}-\epsilon_{\textbf{k},a})^{2}+\frac{\Gamma_{\textbf{k},a}^{2}}{4}}
\end{equation} 
and $\Gamma_{a,\textbf{k}}=1/\tau_{a,\textbf{k}}$ is the inverse of the quasiparticle lifetime. This minimal prescription can be augmented by introducing impurity vertex corrections in one of the spin-flip operators, which restores an exact treatment of disorder in the limit of dilute impurities. 
This task is for the most part beyond the scope of this paper (see next section, however). 
The expression for $\alpha$ in Eq.~(\ref{eq:alpha_beta_SDFT_2}) has already been discussed in a previous paper;\cite{alphaI} hence from here on we shall concentrate on the expression for $\beta$ which now reads
\begin{widetext}
\begin{equation}
\label{eq:beta_0}
\beta^{(0)}=\lim_{q,v_s\to 0}\frac{1}{8\pi s_0}\sum_{a,b}\int_{{\bf k}}|\langle {\bf k+q},b|S^+\,\Delta_{0}({\bf r})|{\bf k},a\rangle|^2 A_a(\epsilon_{\rm F},{\bf k})A_b(\epsilon_{\rm F},{\bf k+q}) \frac{({\bf v}_{{\bf k+q},b}\tau_{{\bf k+q},b}-{\bf v}_{{\bf k},a}\tau_{{\bf k},a})\cdot e{\bf E}}{{\bf q}\cdot{\bf v}_{\rm s}}
\end{equation}
where we have used $\sum_{{\bf k}}\to V\int d^D k/(2\pi)^D\equiv V \int_{{\bf k}}$ with $D$ as the dimensionality, $V$ as the volume and
\begin{equation}
\label{eq:v_s}
{\bf q}\cdot{\bf v}_{\rm s} =\frac{1}{2 s_0}\sum_{a,b}\int_{{\bf k}}|\langle {\bf k+q},b|S^+\,\Delta_{0}({\bf r})|{\bf k},a\rangle|^2\frac{e{\bf v}_{{\bf k+q},b}\tau_{{\bf k+q},b}\delta(\epsilon_{\rm F}-\epsilon_{{\bf k+q},b})-e{\bf v}_{{\bf k},a}\tau_{{\bf k},a}\delta(\epsilon_{\rm F}-\epsilon_{{\bf k},a})}{\epsilon_{{\bf k},a}-\epsilon_{{\bf k+q},b}}.
\end{equation}
In Eq.~(\ref{eq:beta_0}) the superscript ``0'' is to remind of the absence of impurity vertex corrections; . In addition, we recall that $s_0=S_0/V$ is the magnetization of the ferromagnet and $|a{\bf k}\rangle$ is a band eigenstate of the ferromagnet {\em without} disorder. 
It is straightforward to show that Eq.~(\ref{eq:v_s}) reduces to the usual expression $v_s=\sigma_s E/(e s_0)$ for vanishing intrinsic spin-orbit coupling. However, we find that in presence of spin-orbit interaction Eq.~(\ref{eq:v_s}) is no longer connected to the spin conductivity. Determining the precise way in which Eq.~(\ref{eq:v_s}) departs from the conventional formula in real materials is an open problem that may have fundamental and practical repercussions. Expanding the integrand in Eq.~(\ref{eq:beta_0}) to first order in $q$ and rearranging the result we arrive at
\begin{eqnarray}
\label{eq:beta_linearized}
\beta^{(0)}&=&-\frac{1}{8\pi s_0 {\bf q}\cdot{\bf v}_{\rm s}}\sum_{a,b}\int_{\textbf{k}}\left[|\langle a,\textbf{k}|S^{+} \Delta_{0}({\bf r})|b,\textbf{k}\rangle|^{2}+|\langle a,\textbf{k}|S^{-} \Delta_{0}({\bf r}) |b,\textbf{k}\rangle|^{2}\right] A_{a}(\epsilon_{\rm F},\textbf{k})A_{b}^{\prime}(\epsilon_{\rm F},\textbf{k})(\textbf{v}_{\textbf{k},a}\cdot e\textbf{E})(\textbf{v}_{\textbf{k},b}\cdot\textbf{q})\tau_{a}\nonumber\\
&& -\frac{1}{4\pi s_0 {\bf q}\cdot{\bf v}_{\rm
    s}}\sum_{a,b}\int_{\textbf{k}}\mbox{Re }\big[\langle
  b,\textbf{k}|S^{-} \Delta_{0}({\bf r}) \,|a,\textbf{k}\rangle\langle
  a,\textbf{k}|S^{+} \Delta_{0}({\bf r}) \,
  \textbf{q}\cdot\partial_{\textbf{k}}|b,\textbf{k}\rangle 
\nonumber\\
&&~~~~~~~~~~~~~~~~~~~~~~~~~~~~+(S^{+}\leftrightarrow S^{-})\big]
A_{a}(\epsilon_{\rm F},\textbf{k}) A_{b}(\epsilon_{\rm
  F},\textbf{k})(\textbf{v}_{\textbf{k},a}\cdot
e\textbf{E})\tau_{a}\nonumber\\  
\end{eqnarray}
\end{widetext}
where $A'(\epsilon_{\rm F},{\bf k})\equiv 2 (\epsilon_{\rm F}-\epsilon_{{\bf k},a})\Gamma_a/\left[(\epsilon_{\rm F}-\epsilon_{{\bf k},a})^{2}+\Gamma_a^{2}/4\right]^{2}$ stands for the derivative of the spectral function and we have neglected $\partial{\Gamma}/\partial{\textbf{k}}$. 
\begin{figure}
\begin{center}
\scalebox{0.4}{\includegraphics{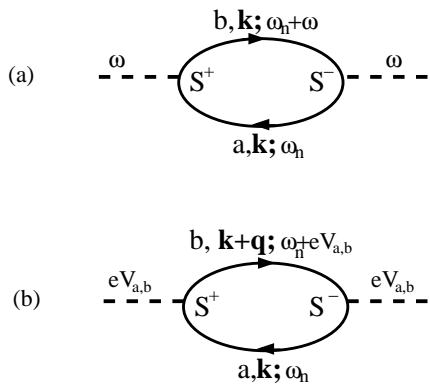}}
\caption{Feynman diagrams for (a) $\alpha$ and (b) $\beta ({\bf q}\cdot{\bf v}_{\rm s})$, the latter with a heuristic consideration of the electric field (for a more rigorous treatment see Appendix A). Solid lines correspond to Green's functions of the band quasiparticles in the Born approximation, dashed lines stand for the magnon of frequency $\omega$ and wavevector {\bf q}, $\omega_n$ is the Matsubara frequency and $e V_{a,b}$ is the difference in the transport deviation energies.  }
\label{fig:ab_bubbles}
\end{center}
\end{figure}
Eq.~(\ref{eq:beta_linearized}) (or Eq.~(\ref{eq:beta_0}))  is the central result of this work and it provides a gateway to evaluate the non-adiabatic STT in materials with complex band structures;\cite{betaII} for a diagrammatic interpretation see Fig.~(\ref{fig:ab_bubbles}). An alternative formula with a similar aspiration has been proposed recently,\cite{tatara3} yet that formula ignores intrinsic spin-orbit interactions and relies on a detailed knowledge of the disorder scattering mechanisms.  In the following three sections we apply Eq.~(\ref{eq:beta_linearized}) to three different simplified models of ferromagnets. For a simpler-to-implement approximate version of Eq.~(\ref{eq:beta_0}) or Eq.~(\ref{eq:beta_linearized}) we refer to Section VI. 

\section{Non-Adiabatic STT for the Parabolic Two-Band Ferromagnet}
The model described in this section bears little resemblance to any real ferromagnet. Yet, it is the only model in which rigorous microscopic results 
for $\beta$ are presently available, thus providing a valuable test bed for Eq.~(\ref{eq:beta_linearized}). The 
mean-field Hamiltonian for itinerant carriers in a two-band Stoner model with parabolic bands is simply
\begin{equation}
H^{(k)}=\frac{k^{2}}{2 m}-\Delta_0 S^{z}
\end{equation}
where $\Delta_0$ is the exchange field and $S^z_{a,b}=\delta_{a,b}\text{sgn}(a)$. 
In this model the eigenstates have no momentum dependence and hence Eq.~(\ref{eq:beta_linearized}) simplifies to
\begin{eqnarray}
\label{eq:stoner_1}
(\textbf{v}_{\rm s}\cdot\textbf{q})\beta^{(0)}
&=&-\frac{\Delta_0^2}{2\pi s_0}\sum_a\int_{\textbf{k}}
A_{a}(\epsilon_{\rm F},\textbf{k}) A_{-a}^{\prime}(\epsilon_{\rm
  F},\textbf{k})
\nonumber\\
&&~~~~~~~~
\frac{\textbf{k}\cdot\textbf{q}}{m}
\frac{\textbf{k}\cdot e\textbf{E}}{m}\tau_{\textbf{k},a}, 
\end{eqnarray}
where $a=+(-)$ for majority (minority) spins, $\textbf{v}_{\textbf{k},\pm}=\textbf{k}/m$, and $S^{\pm}=S^{x}\pm i S^{y}$ with $S^{x}_{a,b}=\delta_{a,b}$. Also, from here on repeated indexes will imply a sum. Taking $\Delta_0\leq E_{\rm F}$ and $\Delta_0>>1/\tau$, the momentum integral in Eq.~(\ref{eq:stoner_1}) is performed in the complex energy plane using a keyhole contour around the branch cut that stems from the 3D density of states:
\begin{widetext}
\begin{eqnarray}
\label{eq:stoner_2}
(\textbf{v}_{\rm s}\cdot\textbf{q})\beta^{(0)} &=&-\frac{\Delta_0^2}{2\pi s_0} \frac{2 e\textbf{E}\cdot\textbf{q}}{3 m}\int_{0}^{\infty} \nu(\epsilon) A_{a}(\epsilon_{{\rm F},a}-\epsilon) A_{-a}^{\prime}(\epsilon_{{\rm F},-a}-\epsilon) \epsilon \tau_{\textbf{k},a}\nonumber\\
&\simeq&\frac{e\textbf{E}\cdot\textbf{q}}{6 m \Delta_0 s_0}\mbox{sgn}(a)\nu_{a}\epsilon_{{\rm F},a}\tau_{a}\Gamma_{-a}\nonumber\\
&=&\frac{e\textbf{E}\cdot\textbf{q}}{2 m \Delta_0 s_0}(n_{\uparrow}\tau_{\uparrow}\gamma_{\downarrow}-n_{\downarrow}\tau_{\downarrow}\gamma_{\uparrow})
\end{eqnarray}
\end{widetext}
where $\epsilon_{{\rm F},a}= \epsilon_{\rm F}+\mbox{sgn}(a) \Delta_0$, $\nu_{a}$ is the spin-dependent density of states at the Fermi surface, $n_{a}=2 \nu_{a}\epsilon_{{\rm F},a}/3$ is the corresponding number density, and $\gamma_{a}\equiv\Gamma_{a}/2$. The factor $1/3$ on the first line of Eq.~(\ref{eq:stoner_2}) comes from the angular integration. 
In the second line of Eq.~(\ref{eq:stoner_2}) we have neglected a term that is smaller than the one retained by a 
factor of $ \Delta_0^{2}/(12 \epsilon_{\rm F}^{2})$; such extra term (which would have been absent in a two-dimensional version of the  model) appears to be missing in previous work.\cite{tatara,rembert}  

The simplicity of this model enables a partial incorporation of impurity vertex corrections. By adding to $\beta^{(0)}$ the contribution from the leading order vertex correction ($\beta^{(1)}$), we shall recover the results obtained previously for this model by a full calculation of the transverse spin response function. As it turns out, $\beta^{(1)}$ is qualitatively 
important because it ensures that only spin-dependent impurities contribute to the non-adiabatic STT in 
the absence of an intrinsic spin-orbit interaction. In Appendix B we derive the following result: 
\begin{widetext}
\begin{equation}
\label{eq:beta_vertex}
(\textbf{v}_{\rm s}\cdot\textbf{q})\beta^{(1)} = \frac{e \Delta_0^2}{4\pi s_0}\int_{\textbf{k},\textbf{k}^{\prime}}u^{i}\mbox{Re }\left[S^{+}_{a,b}S^{i}_{b,b^{\prime}}S^{-}_{b^{\prime},a^{\prime}}S^{i}_{a^{\prime},a}\right]\frac{A_{a}(\epsilon_{\rm F},\textbf{k})}{(\epsilon_{\rm F}-\epsilon_{k^{\prime},a^{\prime}})}
\left[\frac{A_{b}(\epsilon_{\rm F},\textbf{k}+\textbf{q})}{(\epsilon_{\rm F}-\epsilon_{\textbf{k}^{\prime}+\textbf{q},b^{\prime}})} V_{b,a}+ \frac{A_{b^{\prime}}(\epsilon_{\rm F},\textbf{k}^{\prime}+\textbf{q})}{(\epsilon_{\rm F}-\epsilon_{\textbf{k}+\textbf{q},b})}V_{b^{\prime},a}\right],
\end{equation}
where $u^{i}\equiv n_{i} \overline{w_{i}^{2}}$ ($i=0,x,y,z$), $n_{i}$ is the density of scatterers, $w_{i}$ is the Fourier transform of the scattering potential and the overline denotes an average over different disorder configurations.\cite{tatara} Also, $V_{a,b}=(\tau_b v_{{\bf k+q},b}-\tau_a v_{{\bf k},a})\cdot e {\bf E}$. 
Expanding Eq.~(\ref{eq:beta_vertex}) to first order in $q$, we arrive at
\begin{equation}
\label{eq:beta_vertex_2}
(\textbf{v}_{\rm s}\cdot\textbf{q})\beta^{(1)}=-\frac{\Delta_0^2}{2\pi s_0} (u^{0}-u^{z})\int_{\textbf{k},\textbf{k}^{\prime}}\frac{A_{a}(\epsilon_{\rm F},\textbf{k})}{\epsilon_{\rm F}-\epsilon_{\textbf{k}^{\prime},a}}\left[\frac{A_{-a}^{\prime}(\epsilon_{\rm F},\textbf{k})}{\epsilon_{\rm F}-\epsilon_{\textbf{k}^{\prime},-a}}+\frac{A_{-a}(\epsilon_{\rm F},\textbf{k}^{\prime})}{(\epsilon_{\rm F}-\epsilon_{\textbf{k},-a})^{2}}\right]\frac{\textbf{k}\cdot\textbf{q}}{m} \frac{\textbf{k}\cdot e\textbf{E}}{m}\tau_{\textbf{k},a}
\end{equation}
In the derivation of Eq.~(\ref{eq:beta_vertex_2}) we have used $S^{\pm}=S^{x}\pm i S^{y}$ 
and assumed that $u^{x}=u^{y}\equiv u^{x,y}$, so that
$u^{i}\mbox{Re}\left[S^{x}_{a,b}S^{i}_{b,b^{\prime}}S^{x}_{b^{\prime},a^{\prime}}S^{i}_{a^{\prime},a}\right] = \left(u^{0}-u^{z}\right)\delta_{a,a^{\prime}}\delta_{b,b^{\prime}}\delta_{a,-b}$. 
In addition, we have used $\int_{\textbf{k},\textbf{k}^{\prime}} F(|\textbf{k}|,|\textbf{k}^{\prime}|)k_{i} k^{\prime}_{j} =0$. The first term inside the square brackets of Eq.~(\ref{eq:beta_vertex_2}) can be ignored in the weak disorder regime because its contribution is linear in the scattering rate, as opposed to the second term, which contributes at zeroth order. Then,
\begin{eqnarray}
(\textbf{v}_{\rm s}\cdot\textbf{q})\beta^{(1)}&=& -\frac{\Delta_0^2}{\pi s_0}(u^{0}-u^{z})\int_{\textbf{k},\textbf{k}^{\prime}}\frac{A_{a}(\epsilon_{\rm F},\textbf{k})A_{-a}(\epsilon_{\rm F},\textbf{k}^{\prime})}{(\epsilon_{\rm F}-\epsilon_{\textbf{k}^{\prime},a})(\epsilon_{\rm F}-\epsilon_{\textbf{k},-a})^{2}}\frac{\textbf{k}\cdot\textbf{q}}{m} \frac{\textbf{k}\cdot e\textbf{E}}{m}\tau_{\textbf{k},a}\nonumber\\
&\simeq& -\frac{\Delta_0^2}{\pi s_0}(u^{0}-u^{z})\frac{2 e\textbf{E}\cdot\textbf{q}}{3 m}\int_{-\infty}^{\infty} d\epsilon d\epsilon^{\prime}\nu(\epsilon)\nu(\epsilon^{\prime})\frac{A_{a}(\epsilon_{{\rm F},a}-\epsilon) A_{-a}(\epsilon_{{\rm F},-a}-\epsilon^{\prime})}{(\epsilon_{\rm F}-\epsilon^{\prime}_{a})(\epsilon_{\rm F}-\epsilon_{-a})^{2}}\epsilon\tau_{a}\nonumber\\
&\simeq& -\pi(u^{0}-u^{z})\frac{e\textbf{E}\cdot\textbf{q}}{2 m \Delta_0 s_0}\mbox{sign}(a) n_{a}\tau_{a}\nu_{-a}
\end{eqnarray} 
Combining this with Eq.~(\ref{eq:stoner_2}), we get
\begin{eqnarray}
\label{eq:beta_3D}
(\textbf{v}_{\rm s}\cdot\textbf{q})\beta & \simeq& (\textbf{v}_{\rm s}\cdot\textbf{q})\beta^{(0)}+(\textbf{v}_{\rm s}\cdot\textbf{q})\beta^{(1)}\nonumber\\
&=& \frac{e\textbf{E}\cdot\textbf{q}}{2 m s_0 \Delta_0}\left[n_{\uparrow}\tau_{\uparrow}\gamma_{\downarrow}-n_{\downarrow}\tau_{\downarrow}\gamma_{\uparrow}-\pi(u^{0}-u^{z})(n_{\uparrow}\tau_{\uparrow}\nu_{\downarrow}-n_{\downarrow}\tau_{\downarrow}\nu_{\uparrow})\right]\nonumber\\
&=& \pi\frac{e\textbf{E}\cdot\textbf{q}}{m s_0 \Delta_0}\left[n_{\uparrow}\tau_{\uparrow}\left(u^{z}\nu_{\downarrow}+u^{x,y}\nu_{\uparrow}\right)-n_{\downarrow}\tau_{\downarrow}\left(u^{z}\nu_{\uparrow}+u^{x,y}\nu_{\downarrow}\right)\right]
\end{eqnarray}
\end{widetext}
where we have used $\gamma_{a}=\pi\left[(u^{0}+u^{z})\nu_{a}+2 u^{x,y}\nu_{-a}\right]$. 
In this model it is simple to solve Eq.~(\ref{eq:v_s}) for ${\bf v}_{\rm s}$ analytically, whereupon Eq.~(\ref{eq:beta_3D}) agrees with the results published by other authors in Refs.[~\onlinecite{tatara,rembert}] from full non-linear response function calculations. However, we reiterate that in order to reach such agreement we had to neglect a term of order $\Delta_0^2/\epsilon_{\rm F}^2$ in Eq.~(\ref{eq:stoner_2}). This extra term is insignificant in all but nearly half metallic ferromagnets.

\section{Non-Adiabatic STT for a Magnetized Two-Dimensional Electron Gas}

The model studied in the previous section misses the intrinsic spin-orbit interaction that is inevitably present in the band structure of actual ferromagnets. Furthermore, since intrinsic spin-orbit interaction is instrumental for the Gilbert damping at low temperatures, a similarly prominent role may be expected in regards to the non-adiabatic spin transfer torque. Hence, the present section is devoted to investigate the relatively unexplored\cite{tatara2,tatara3} effect of intrinsic spin-orbit interaction on $\beta$. The minimal model for this enterprise is the two-dimensional electron-gas ferromagnet with Rashba spin-orbit interaction, represented by
\begin{equation}
H^{(k)}=\frac{k^{2}}{2 m}-\textbf{b}\cdot\textbf{S},
\end{equation}
where $\textbf{b}=(\lambda k_{y},-\lambda k_{x},\Delta_0)$, $\lambda$ is the Rashba spin-orbit coupling strength and $\Delta_0$ is the exchange field.

\begin{figure}
\begin{center}
\scalebox{0.4}{\includegraphics{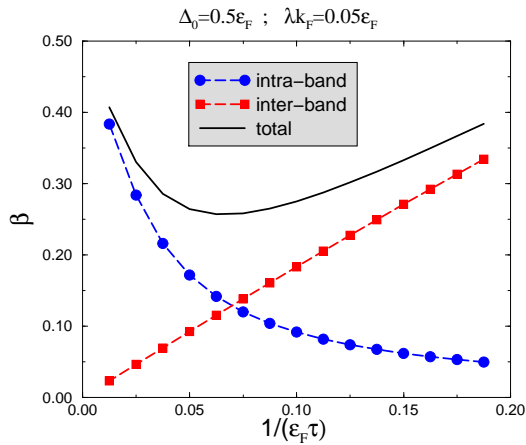}}
\caption{\textbf{M2DEG}: inter-band contribution, intra-band
  contribution and the total non-adiabatic STT for a magnetized
  two-dimensional electron gas (M2DEG). In this figure the exchange field dominates over the spin-orbit splitting. At higher disorder the inter-band part (proportional to resistivity) dominates, while at low disorder the inter-band part (proportional to conductivity) overtakes. For simplicity, the scattering time $\tau$ is taken to be the same for all sub-bands. }
\label{fig:M2DEG}
\end{center}
\end{figure}

The eigenspinors of this model are $|+,\textbf{k}\rangle=\left(\cos(\theta/2), -i \exp(i\phi)\sin(\theta/2)\right)$ and $|-,\textbf{k}\rangle=\left(\sin(\theta/2),i \exp(i\phi)\cos(\theta/2)\right)$, where the spinor angles are defined through $\cos\theta=\Delta_0/\sqrt{\lambda^{2}k^{2}+\Delta_0^{2}}$ and $\tan\phi=k_y/k_x$. The corresponding eigenenergies are $E_{\textbf{k}\pm}=k^{2}/(2 m)\mp\sqrt{\Delta_0^{2}+\lambda^{2}k^{2}}$. Therefore, the band velocities are given by $\textbf{v}_{\textbf{k}\pm}=\textbf{k}\left(1/m\mp\lambda^{2}/\sqrt{\lambda^{2}k^{2}+\Delta_0^{2}}\right)={\bf k}/m_{\pm}$.
Disregarding the vertex corrections, the non-adiabatic spin-torque of this model 
may be evaluated analytically starting from Eq.~(\ref{eq:beta_linearized}).  
We find that (see Appendix C):
\begin{widetext}
\begin{eqnarray}
\label{eq:beta_M2DEG}
(\textbf{v}_{\rm s}\cdot\textbf{q})\beta^{(0)}&\simeq & \frac{\Delta_0^2 e\textbf{E}\cdot\textbf{q}}{8 \pi s_0}\left[\frac{m^{2}}{4 m_{+}m_{-}}\left(1+\frac{\Delta_0^{2}}{b^{2}}\right)\frac{1}{b^{2}}+\frac{1}{4}\frac{\lambda^{2}k_{\rm F}^{2}\Delta_0^{2}}{b^{6}}\right]\nonumber\\
&+&\frac{\Delta_0^2 e\textbf{E}\cdot\textbf{q}}{8 \pi s_0}\left[\frac{1}{2}\frac{m^{2}}{m_{+}^{2}}\frac{\lambda^{2}k_{\rm F}^{2}}{b^{2}}\left(1-\frac{\delta m_{+}}{m}\frac{\Delta_0^{2}}{b^{2}}\right)\tau^{2}+\frac{1}{2}\frac{m^{2}}{m_{-}^{2}}\frac{\lambda^{2}k_{\rm F}^{2}}{b^{2}}\left(1-\frac{\delta m_{-}}{m}\frac{\Delta_0^{2}}{b^{2}}\right)\tau^{2}\right]
\end{eqnarray}
\end{widetext}
where $b=\sqrt{\lambda^{2}k_{\rm F}^{2}+\Delta_0^{2}}$ ($k_{\rm F}=\sqrt{2 m \epsilon_{\rm F}}$), and $\delta m_{\pm}=m-m_{\pm}$ . As we explain in the Appendix, Eq.~(\ref{eq:beta_M2DEG}) applies for $\lambda k_{\rm F}, \Delta_0, 1/\tau <<\epsilon_{\rm F}$; for a more general analysis,  Eq.~(\ref{eq:beta_linearized}) must be solved numerically (e.g. see Fig.~(\ref{fig:M2DEG})). Eq.~(\ref{eq:beta_M2DEG}) reveals that intrinsic spin-orbit interaction enables \emph{intra-band} contributions to $\beta$, whose signature is the $O(\tau^{2})$ dependence on the second line. In contrast, the \emph{inter-band} contributions appear as $O(\tau^{0})$.
Since $\textbf{v}_{\rm s}$ itself is linear in the scattering time, it follows that $\beta$ is proportional to the electrical conductivity in the clean regime and the resistivity in the disordered regime, much like the Gilbert damping $\alpha$. We expect this qualitative feature to be model-independent and applicable to real ferromagnets. 

\section{Non-Adiabatic STT for $\text{(Ga,Mn)As}$}
In this section we shall apply Eq.~(\ref{eq:beta_linearized}) to a more sophisticated model which provides a reasonable description of (III,Mn)V magnetic semiconductors.\cite{gamnas} Since the orbitals at the Fermi energy are very similar to the states near the top of the valence band of the host (III,V) semiconductor, the electronic structure of (III,Mn)V ferromagnets is remarkably simple. Using a p-d mean field theory model for the ferromagnetic ground state and a four-band spherical model for the host semiconductor band structure, $\text{Ga}_{1-x} \text{Mn}_{x} \text{As}$ may be described by 
\begin{equation}
\label{eq:H_GaMnAs}
H^{(k)}=\frac{1}{2 m}\left[\left(\gamma_{1}+\frac{5}{2}\gamma_{2}\right)k^{2}-2\gamma_{3}(\textbf{k}\cdot\textbf{S})^{2}\right]+\Delta_0 S_{z},
\end{equation}
where $\textbf{S}$ is the spin operator projected onto the J=3/2 total angular momentum subspace at the top of the valence band and \{$\gamma_{1}=6.98, \gamma_{2}=\gamma_{3}=2.5$\} are the Luttinger parameters for the spherical approximation to the 
valence bands of GaAs. In addition, $\Delta_0=J_{\rm pd} s N_{\rm Mn}=J_{\rm pd} s_0$ is the exchange field, $J_{\rm pd}=55 \mbox{ meV} \mbox{nm}^{3}$ is the p-d exchange coupling, $s=5/2$ is the spin of Mn ions, $N_{\rm Mn}=4x/a^{3}$ is the density of Mn ions and $a=0.565 \mbox{ nm}$ is the lattice constant of GaAs. We solve Eq.~(\ref{eq:H_GaMnAs}) numerically and input the outcome in Eqs.~(\ref{eq:v_s}),~(\ref{eq:beta_linearized}). 
\begin{figure}
\begin{center}
\scalebox{0.4}{\includegraphics{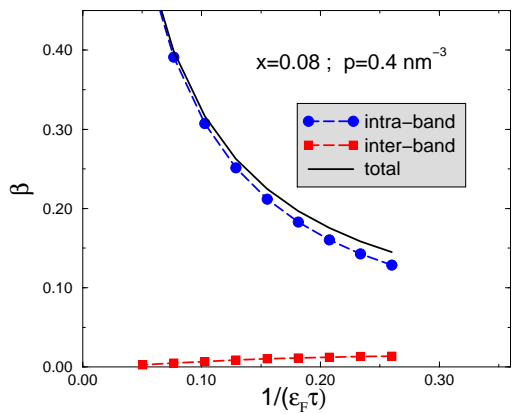}}
\caption{\textbf{GaMnAs}: $\beta^{(0)}$ for $\textbf{E}$ perpendicular to the easy axis of magnetization ($\hat z$). $x$ and $p$ are the Mn fraction and the hole density, respectively. The intra-band contribution is considerably larger than the inter-band contribution, due to the strong intrinsic spin-orbit interaction. Since the 4-band model typically overestimates the influence of intrinsic spin-orbit interaction, it is likely that the dominion of intra-band contributions be reduced in the more accurate 6-band model. By evaluating $\beta$ for ${\bf E}||\hat z$ (not shown) we infer that it does not depend significantly on the relative direction between the magnetic easy axis and the electric field. }
\label{fig:GaMnAs_perp}
\end{center}
\end{figure}

The results are summarized in Fig.~(\ref{fig:GaMnAs_perp}). We find that the intra-band contribution dominates as a consequence of the strong intrinsic spin-orbit interaction, much like for the Gilbert damping;\cite{alphaII}. Incidentally, $\beta$ barely changes regardless of whether the applied electric field is along the easy axis of the magnetization or perpendicular to it.

\section{$\alpha/\beta$ in real materials}

The preceding three sections have been focused on testing and analyzing Eq.~(\ref{eq:beta_linearized}) for specific models of ferromagnets.  In this section we return to more general considerations and survey the phenomenologically important quantitative relationship between $\alpha$ and $\beta$ in realistic ferromagnets,
which always have intrinsic spin-orbit interactions. We begin by recollecting the expression for the Gilbert damping coefficient derived elsewhere:\cite{alphaI}
\begin{equation}
\label{eq:alpha_recall}
\alpha=\frac{1}{8\pi s_0}\sum_{a,b}\int_{\textbf{k}}|\langle b,\textbf{k}|S^{+}\Delta_0|a,\textbf{k}\rangle|^{2} A_{a}(\epsilon_{\rm F},\textbf{k})A_{b}(\epsilon_{\rm F},\textbf{k})
\end{equation}
where we have ignored disorder vertex corrections. This expression is to be compared with Eq.~(\ref{eq:beta_0}); for pedagogical purposes we discuss 
intra-band and inter-band contributions separately.

Starting from Eq.~(\ref{eq:beta_0}) and expanding the integrand to first order in ${\bf q}$ we obtain
\begin{eqnarray}
\label{eq:beta_intra}
\beta_{intra}&=&\frac{1}{8\pi s_0}\int_{\textbf{k}}|\langle
a,\textbf{k}|S^{+}\Delta_0|a,\textbf{k}\rangle|^{2 
} A_{a}(\epsilon_{\rm F},\textbf{k})^2 
\nonumber\\
&&~~~~~~~~~~\frac{e \tau_{a}
  q^{i}\partial_{k_{i}}v_{\textbf{k},a}^{j} E^{j}}{{\bf q}\cdot{\bf
    v}_{\rm s}} 
\end{eqnarray}
 where we have neglected the momentum dependence of the scattering lifetime and a sum over repeated indices is implied. Remarkably, only matrix elements that are diagonal in momentum space contribute to $\beta_{\rm intra}$ ; the implications of this will be highlighted in the next section. Recognizing that $\partial_{k_{j}}v^{i}_{k,a}=(\textbf{1/m})_{a}^{i,j}$, where $(1/\textbf{m})_{a}$ is the inverse effective mass tensor corresponding to band $a$, Eq.~(\ref{eq:beta_intra}) can be rewritten as
\begin{equation}
\label{eq:intra_pre}
\beta_{\rm intra}=\frac{1}{8\pi s_0}\int_{\textbf{k}}|\langle a,\textbf{k}|S^{+}\Delta_0|a,\textbf{k}\rangle|^{2
} A_{a}(\epsilon_{\rm F},\textbf{k})^2\,\frac{{\bf q}\cdot {\bf v}_{{\rm d},a}}{{\bf q}\cdot {\bf {v}_{\rm s}}},
\end{equation}
where 
\begin{equation}
{\bf v}_{{\rm d},a}^i=e\tau_{a}({\bf m}^{-1})_a^{i,j}\textbf{E}^j
\end{equation}
 is the ``drift velocity'' corresponding to the quasiparticles in  band $a$. For Galilean invariant systems\cite{barnes} $v_{{\rm d},a}=v_s$ for any $({\bf k},a)$ and consequently $\beta_{\rm intra}=\alpha_{\rm intra}$.
At first glance, it might appear that $v_s$, which (at least in absence of spin-orbit interaction) is determined by the spin current, must be different than $v_{{\rm d},a}$.  However, recall that $v_s$ is determined by the ratio of the spin current to the magnetization.  If the same electrons contribute to the transport as to the magnetization, $v_s=v_{{\rm d},a}$ provided the scattering rates and the masses are the same for all states.  These conditions are the conditions for an electron system to be Galilean invariant.
\begin{widetext}
The interband contribution can be simplified by noting that  
\begin{equation}
\tau_{b}v^{i}_{\textbf{k}+\textbf{q},b}-\tau_{a} v^{i}_{\textbf{k},a}=(\tau_{b}v^{i}_{\textbf{k}+\textbf{q},b}-\tau_{a} v^{i}_{\textbf{k}+\textbf{q},a})+(\tau_{a}v^{i}_{\textbf{k}+\textbf{q},a}-\tau_{a} v^{i}
_{\textbf{k},a}).
\label{eq:32}
\end{equation}
The second term on the right hand side of Eq.(~\ref{eq:32}) can then be manipulated exactly as in the intra-band case to arrive at
\begin{equation}
\beta_{\rm inter}=\frac{1}{8\pi s_0}\sum_{a,b (a\neq b)}\int_{\textbf{k}}|\langle b,\textbf{k}|S^{+}\Delta_0|a,\textbf{k}\rangle|^{2} A_{a}(\epsilon_{\rm F},\textbf{k})A_{b}(\epsilon_{\rm F},\textbf{k}) \frac{{\bf q}\cdot {\bf v}_{{\rm d},a}}{{\bf q}\cdot {\bf {v}_{\rm s}}}+\delta\beta_{\rm inter}
\end{equation}
where
\begin{equation}
\label{eq:beta_inter}
\delta\beta_{\rm inter}
=\frac{1}{8\pi s_0}\sum_{a,b (a\neq b)}\int_{\textbf{k}}|\langle a,\textbf{k}-\textbf{q}|S^{+}\Delta_0|b,\textbf{k}\rangle|^{2
} A_{a}(\epsilon_{\rm F},\textbf{k}-\textbf{q})A_{b}(\epsilon_{\rm F},\textbf{k}) \frac{(\tau_{b}\textbf{v}_{\textbf{k},b}- \tau_{a}\textbf{v}_{\textbf{k},a})\cdot{\bf E}}{{\bf q}\cdot{\bf v}_{\rm s}}.
\end{equation}
\end{widetext}
When Galilean invariance is preserved the quasiparticle velocity and scattering times are the same for all bands, which implies that $\delta\beta=0$ and 
hence that $\beta_{\rm inter}=\alpha_{\rm inter}$. Although realistic materials are not Galilean invariant, $\delta\beta$ is
nevertheless probably not significant because the term between parenthesis in Eq.~(\ref{eq:beta_inter}) has an oscillatory behavior prone to cancellation. 
The degree of such cancellation must ultimately be determined by realistic calculations for particular materials.

With this proviso, we estimate that
\begin{eqnarray}
\label{eq:beta_approx}
\beta&\simeq&\frac{1}{8\pi s_0}\int_{\textbf{k}}|\langle
b,\textbf{k}|S^{+}\Delta_0|a,\textbf{k}\rangle|^{2}
A_{a}(\epsilon_{\rm F},\textbf{k})A_{b}(\epsilon_{\rm F},\textbf{k})
\nonumber\\
&&~~~~~~~~~~~~~~~~~
\frac{{\bf q}\cdot {\bf v}_{{\rm d},a}}{{\bf q}\cdot {\bf {v}_{\rm
      s}}}. 
\end{eqnarray}
As long as $\delta\beta\simeq 0$ is justified, the simplicity of Eq.~(\ref{eq:beta_approx}) in comparison to Eq.~(\ref{eq:beta_0}) or~(\ref{eq:beta_linearized}) makes of the former the preferred starting point for electronic structure calculations. 
Even when $\delta\beta\neq 0$ Eq.~(\ref{eq:beta_approx}) may be an adequate platform for {\em ab-initio} studies on weakly disordered transition metal ferromagnets and strongly spin-orbit coupled ferromagnetic semiconductors,\cite{caveat} where $\beta$ is largely determined by the intra-band contribution.  
 Furthermore, a direct comparison between Eq.~(\ref{eq:alpha_recall}) and Eq.~(\ref{eq:beta_approx}) leads to the following observations.
  First, for nearly parabolic bands with nearly identical curvature,
  where the ``drift velocity'' is weakly dependent on momentum or the
  band index, we obtain $\beta\simeq (v_{\rm d}/v_{\rm s}) \alpha$ and thus $\beta/\alpha$ is roughly proportional to the ratio of the total spin density to the itinerant spin density, in concordance with predictions from toy models.\cite{tserkovnyak}
Second, if $\alpha/\beta>0$ for a system with purely electron-like
carriers, then $\alpha/\beta>0$ for the same system with purely
hole-like carriers because for a fixed carrier polarization $v_{\rm
  d}^a$ {\em and} $v_{\rm s}$ reverse their signs under $m\to -m$. However, if both hole-like and electron-like carriers coexist at the Fermi energy, then the integrand in Eq.~(\ref{eq:beta_approx}) is positive for some values of $a$ and negative for others. In such situation it is conceivable that  $\alpha/\beta$ be either positive or negative. A negative value of $\beta$ implies a {\em decrease} in magnetization damping due to an applied current. 

\begin{figure}
\begin{center}
\scalebox{0.4}{\includegraphics{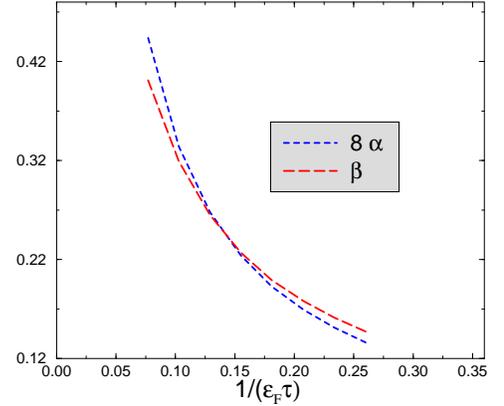}}
\caption{Comparison of $\alpha$ and $\beta$ in (Ga,Mn)As for $x=0.08$ and $p=0.4 nm^{-3}$. It follows that $\beta/\alpha\simeq 8$, with a weak dependence on the scattering rate off impurities. If we use the torque correlation formula (Section VII), we obtain $\beta/\alpha\simeq 10$.} 
\label{fig:alpha_beta}
\end{center}
\end{figure}

As an illustration of the foregoing discussion, in Fig.~(\ref{fig:alpha_beta}) we evaluate $\alpha/\beta$ for (Ga,Mn)As. We find $\beta$ to be about an order of magnitude larger than $\alpha$, which is reasonable because  
(i) the local moment magnetization is larger than the valence band hole magnetization, and (ii) the spin-orbit coupling in the valence band decreases the transport spin polarization. Accordingly $\beta$ is of the order of unity, in qualitative agreement with recent theoretical work\cite{brataas}.

\section{Torque-Correlation Formula for the Non-Adiabatic STT}
Thus far we have evaluated non-adiabatic STT using the bare vertex
$\langle a,\textbf{k}|S^{+}|b,\textbf{k+q}\rangle$. In this section,
we shall analyze an alternative matrix element denoted $\langle
a,\textbf{k}|K|b,\textbf{k+q}\rangle$ (see below for an explicit
expression),  which may be better suited to realistic electronic
structure calculations.\cite{kambersky,betaII} 
We begin by making the approximation that the exchange splitting can
be written as a constant spin-dependent shift $H_{\rm ex}=\Delta_0
S^{z}$.  Then, the mean-field
quasiparticle Hamiltonian 
$H^{(k)}=H_{\rm kin}^{(k)}+H_{\rm so}^{(k)}+H_{\rm ex}$
can be written as the sum of a 
spin-independent part $H_{\rm kin}^{(k)}$, the exchange term, and the
spin-orbit coupling $H_{\rm so}^{(k)}$.  With this approximation, we
have the identity:
\begin{eqnarray}
\label{eq:n_vs_k}
&&\langle a,\textbf{k}|S^{+}|b,\textbf{k}+\textbf{q}\rangle\nonumber\nonumber\\
&=& \frac{1}{\Delta_0}\langle a,\textbf{k}|\left[H^{(k)},S^{+}\right]|b,\textbf{k}+\textbf{q}\rangle\nonumber\\
&-&\frac{1}{\Delta_0}\langle a,\textbf{k}|\left[H_{\rm so}^{(k)},S^{+}\right]|b,\textbf{k}+\textbf{q}\rangle .
\end{eqnarray}
The last term in
the right hand side of Eq.~(\ref{eq:n_vs_k}) is the generalization of
the torque matrix element used in \emph{ab-initio} calculations of the
Gilbert damping:
\begin{equation}
\label{eq:kambersky}
\langle a,\textbf{k}|K|b,\textbf{k}+\textbf{q}\rangle\equiv\frac{1}{\Delta_0}\langle a,\textbf{k}|\left[H_{\rm so}^{(k)},S^{+}\right]|b,\textbf{k}+\textbf{q}\rangle 
\end{equation} 

\begin{figure}
\begin{center}
\scalebox{0.4}{\includegraphics{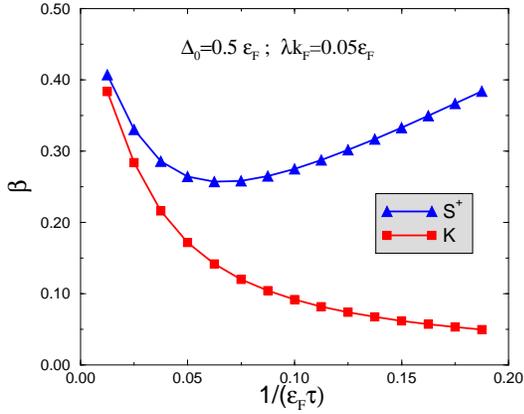}}
\caption{\textbf{M2DEG}: comparing $S$ and $K$ matrix element expressions for the 
non-adiabatic STT formula in the weakly spin-orbit coupled regime. Both formulations agree in the clean limit, where the intra-band contribution is dominant. In more disordered samples inter-band contributions become more visible and $S$ and $K$ begin to differ; the latter is known to be more accurate in the weakly spin-orbit coupled regime.}
\label{fig:K vs S M2DEG 1}
\end{center}
\end{figure}

\begin{figure}
\begin{center}
\scalebox{0.4}{\includegraphics{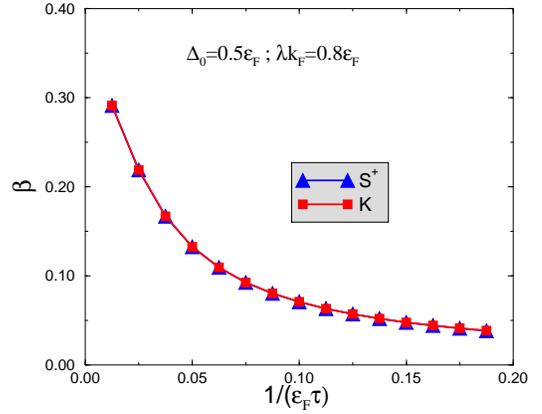}}
\caption{\textbf{M2DEG}: In the strongly spin-orbit coupled limit the intra-band contribution reigns over the inter-band contribution and accordingly $S$ and  $K$ matrix element expressions display a good (excellent in this figure) agreement. Nevertheless, this agreement does not guarantee quantitative reliability, because for strong spin-orbit interactions impurity vertex corrections may play an important role.} 
\label{fig:K vs S M2DEG 2}
\end{center}
\end{figure}

Eq.~(\ref{eq:n_vs_k}) implies that at 
${\bf q=0}$ $\langle b,{\bf k}|S^{+}|a,{\bf k}\rangle\simeq\langle b,{\bf k}|K|a,{\bf k}\rangle$ provided that $(E_{\textbf{k},a}-E_{\textbf{k},b})<<\Delta_0$, 
which is trivially satisfied for intra-band transitions but less so for inter-band transitions.\cite{alphaII} For ${\bf q\neq 0}$ the agreement between intra-band matrix elements is no longer obvious and is affected by the momentum dependence of the band eigenstates. At any rate, Eq.~(\ref{eq:beta_intra}) demonstrates that only ${\bf q=0}$ matrix elements contribute to $\beta_{\rm intra}$; therefore $\beta_{\rm intra}$ has the same value for $S$ and $K$ matrix elements. The disparity between the two formulations is restricted to $\beta_{\rm inter}$, and may be significant if the most prominent inter-band matrix elements connect states that are {\em not} close in energy. When they disagree, it is generally
unclear\cite{exact} whether $S$ or $K$ matrix elements will yield a better estimate of $\beta_{\rm inter}$. The weak spin-orbit limit is a possible exception, in which the use of $K$ appears to offer a practical advantage over $S$. In this regime $S$ generates a spurious inter-band contribution in 
the absence of magnetic impurities (recall Section III) and it is only after the inclusion of the leading order vertex correction that such deficiency gets remedied. In contrast, $K$ vanishes identically in absence of spin-orbit interactions, thus bypassing the pertinent problem without having to introduce vertex corrections. 

\begin{figure}
\begin{center}
\scalebox{0.4}{\includegraphics{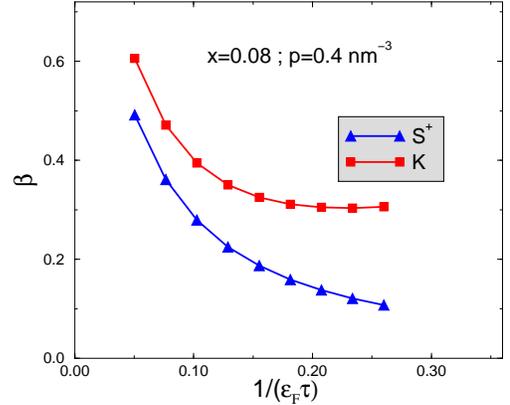}}
\caption{\textbf{GaMnAs}: comparison between $S$ and $K$ matrix element expressions for ${\bf E}\perp\hat z$. The disagreement between both formulations stems from inter-band transitions, which are less important as $\tau$ increases. Little changes when ${\bf E}\parallel\hat z$.}
\label{fig:K vs S GaMnAs}
\end{center}
\end{figure}

Figs.~(\ref{fig:K vs S M2DEG 1})-~(\ref{fig:K vs S GaMnAs}) display a quantitative comparison between the non-adiabatic STT obtained from $K$ and $S$, both for the M2DEG and (Ga,Mn)As. Fig.~(\ref{fig:K vs S M2DEG 1}) reflects the aforementioned overestimation of $S$ in the inter-band dominated regime of weakly spin-orbit coupled ferromagnets. In the strong spin-orbit limit, where intra-band contributions dominate in the disorder range of interest,  $K$ and $S$ agree fairly well (Figs.~(\ref{fig:K vs S M2DEG 2}) and~(\ref{fig:K vs S GaMnAs})). Summing up, insofar as impurity vertex corrections play a minor role and the dominant contribution to $\beta$ stems from intra-band transitions the torque-correlation formula will provide a reliable estimate of $\beta$. 

\section{Connection to the Effective Field Model}

As explained in Section II we view the non-adiabatic STT as the change in magnetization damping due to a transport current. 
The present section is designed to complement that understanding from a different perspective based on an effective field formulation, which provides a simple physical interpretation for both intra-band and inter-band contributions to $\beta$. 

An effective field ${\cal H}^{\rm eff}$ may be expressed
as the variation of the system energy with respect to the magnetization direction ${\cal H}^{\rm eff}_{i}
= -(1/s_0) \partial E / \partial \Omega_{i}$.  Here we approximate the energy with the Kohn-Sham
eigenvalue sum
~
\begin{equation}
E = \sum_{{\bf k},a} n_{{\bf k},a} \epsilon_{{\bf k},a} \, .
\end{equation}

\noindent The variation of this energy with respect to the magnetization direction yields
~
\begin{equation}
\label{eq:h_eff_total}
{\cal H}^{\rm eff}_{i} = -\frac{1}{s_0} \sum_{{\bf k},a} \left [ n_{{\bf k},a} \frac{\partial \epsilon_{{\bf
k},a}}{\partial \Omega_{i}} + \frac{\partial n_{{\bf k},a}}{\partial \Omega_{i}} \epsilon_{{\bf
k},a} \right ] \, .
\end{equation}

\noindent It has previously been shown that, in the absence of current, the first term in the
sum leads to intra-band Gilbert damping\cite{breathing,fahnle}  while the second term produces inter-band damping.\cite{keith}  In the 
following, we generalize these results by allowing the flow of an electrical current.  $\alpha$ and $\beta$ may be extracted by identifying the the dissipative part of
the effective field 
with $-\alpha \partial \hat{\Omega}/\partial t - \beta {\bf v}_{\rm s}\cdot \nabla \hat{\Omega}$ that appears in the LLS equation.

{\em Intra-band terms}: We begin by recognizing that as the direction of magnetization changes in 
time, so does the shape of the Fermi surface, provided that there is an intrinsic spin-orbit 
interaction.  Consequently, empty (full) states appear below (above) the Fermi energy, giving 
rise to an out-of-equilibrium quasiparticle distribution. 
This configuration tends to relax back to equilibrium, but repopulation requires a time  
$\tau$.  Due to the time delay, the quasiparticle distribution lags behind the 
dynamical configuration of the Fermi surface, effectively creating a friction (damping) force 
on the magnetization.  From a quantitative standpoint, the preceding discussion means that the 
quasiparticle energies $\epsilon_{{\bf k}, a}$ follow the magnetization adiabatically, whereas 
the occupation numbers $n_{{\bf k}, a}$ deviate from the instantaneous equilibrium distribution 
$f_{{\bf k},a}$ via
~
\begin{equation}
\label{eq:n}
n_{{\bf k}, a}=f_{{\bf k}, a}-\tau_{{\bf k},a}\left(\frac{\partial f_{{\bf k}, a}}{\partial t}+\dot{{\bf r}_a}\cdot\frac{\partial f_{{\bf k}, a}}{\partial {\bf r}}+\dot{{\bf k}}\cdot\frac{\partial f_{{\bf k}, a}}{\partial {\bf k}}\right),
\end{equation}

\noindent where we have used the relaxation time approximation. As we explain below, the last two terms 
in Eq.~(\ref{eq:n}) do not contribute to damping in the absence of an electric field and have thus 
been ignored by prior applications of the breathing Fermi surface model, which concentrate on 
Gilbert damping. 
It is customary to associate intra-band magnetization damping with the torque exerted by the 
part of the effective field 
~
\begin{equation}
\label{eq:h_eff_bis}
{\cal H}^{\rm eff}_{\rm intra} = -\frac{1}{s_0}\sum_{{\bf k},a} n_{{\bf k},a} \frac{\partial \epsilon_{{\bf 
k},a}}{\partial \hat\Omega}
\end{equation}

\noindent that is lagging behind the instantaneous magnetization.  Plugging Eq.~(\ref{eq:n}) 
in Eq.~(\ref{eq:h_eff_bis}) we obtain
~
\begin{widetext}
\begin{equation}
\label{eq:h_eff_full}
{\cal H}^{\rm eff}_{{\rm intra, i}}=\frac{1}{s_0}\sum_{{\bf k},a}\left[ -f_{{\bf k}, a}\frac{\partial\epsilon_{{\bf k}, a}}{\partial\Omega_i}+
\tau_a\frac{\partial{f_{{\bf k}, a}}}{\partial\epsilon_{{\bf k}, a}}\frac{\partial\epsilon_{{\bf k}, a}}{\partial\Omega_i}\frac{\partial\epsilon_{{\bf k}, a}}{\partial\Omega_j}\frac{\partial\Omega_j}{\partial t}+
\tau_a \dot{r}_a^l\frac{\partial{f_{{\bf k}, a}}}{\partial\epsilon_{{\bf k}, a}}\frac{\partial\epsilon_{{\bf k}, a}}{\partial\Omega_i}\frac{\partial\epsilon_{{\bf k}, a}}{\partial\Omega_j}\frac{\partial\Omega_j}{\partial {r_l}}+
\tau_a \dot{k}^j\frac{\partial{f_{{\bf k}, a}}}{\partial\epsilon_{{\bf k}, a}}\frac{\partial\epsilon_{{\bf k}, a}}{\partial k_j}\frac{\partial\epsilon_{{\bf k}, a}}{\partial\Omega_i}\right]
\end{equation}
\end{widetext}

\noindent where a sum is implied over repeated Latin indices. 
The first term in Eq.~(\ref{eq:h_eff_full}) is a contribution to the anisotropy field; it 
evolves in synchrony 
with the dynamical Fermi surface and is thus the reactive component of the effective field. The 
remaining terms, which describe the time lag of the effective field due to a nonzero relaxation 
time, are responsible for intra-band damping. 
The last term vanishes in crystals with inversion symmetry because $\dot{k}=e E$ and
$\partial\epsilon/\partial{\bf k}$ is an odd function of momentum. 
Similarly, if we take $\dot{{\bf r}}=\partial\epsilon({\bf k})/\partial {\bf k}$ the second to 
last term ought to vanish as well. This leaves us with the first two terms in Eq. 
(~\ref{eq:h_eff_full}), which capture the intra-band Gilbert damping but not the non-adiabatic 
STT. 
This is not surprising as the latter involves the {\em coupled} response to spatial variations 
of magnetization and a weak electric field, rendering linear order in perturbation theory 
insufficient (see Appendix A). 
In order to account for the relevant non-linearity we use $\dot{{\bf r}}=\partial\epsilon({\bf 
k}-e {\bf v}\cdot{\bf E}\tau)/\partial {\bf k}$ in Eq.(~\ref{eq:h_eff_full}), where ${\bf 
v}=\partial\epsilon({\bf k})/\partial {\bf k}$. The dissipative part of ${\cal H}^{\rm 
eff}_{\rm intra}$ then reads
~
\begin{equation}
\label{eq:h_damp}
{\cal H}^{\rm eff,damp}_{\rm intra, i}=\frac{1}{s_0}\sum_{{\bf k},a} \tau_{{\bf k}, a}\frac{\partial{f_{{\epsilon_{\bf k}}, 
a}}}{\partial\epsilon_{{\bf k},a}}\frac{\partial\epsilon_{{\bf k}, a}}{\partial\Omega_i}\frac{\partial\epsilon_{{\bf k}, a}}{\partial\Omega_j}\left[\frac{\partial\Omega_j}{\partial t}
+{\bf v}_{{\rm d},a}^l\frac{\partial\Omega_j}{\partial r^l}\right],
\end{equation}

\noindent where ${\bf v}_{d,a}^i=e\tau_{a}(m^{-1})_a^{i,j}\textbf{E}^j$ is the ``drift 
velocity'' corresponding to band $a$.
Eq.~(\ref{eq:h_damp}) may now be identified with $-\alpha_{\rm intra}\partial\hat\Omega/\partial 
t-\beta_{\rm intra}{\bf v}_s\cdot\nabla\hat\Omega$ that appears in the LLS equation. For an isotropic 
system this results in
~
\begin{eqnarray}
\label{eq:a_b_breathe}
\alpha_{\rm intra}&=&-\frac{1}{s_0}\sum_{{\bf k}, a, i}\tau_{{\bf k}, a}\frac{\partial f_{{\bf 
k}, 
a}}{\partial\epsilon_{{\bf k}, a}}\left(\frac{\partial\epsilon_{{\bf k},a}}{\partial\Omega_i}\right)^2\nonumber\\
\beta_{\rm intra}&=&-\frac{1}{s_0}\sum_{{\bf k}, a, i}\tau_{{\bf k}, a}\frac{\partial f_{{\bf 
k}, 
a}}{\partial\epsilon_{{\bf k}, a}}\left(\frac{\partial\epsilon_{{\bf k}, a}}{\partial\Omega_i}\right)^2\frac{{\bf q}\cdot{{\bf v}_{d,a}}}{{\bf q}\cdot{\bf v}_s}.
\end{eqnarray}

\noindent Since $\langle [S^{x},H_{so}]\rangle=\partial_{\phi}\langle \exp(i S^{x}\phi) H_{so} 
\exp(-i 
S^{x}\phi)\rangle=\partial\epsilon/\partial\phi$ for an infinitesimal angle of rotation $\phi$ around the instantaneous magnetization, $\beta$ in Eq.~(\ref{eq:a_b_breathe}) may be rewritten as
~
\begin{equation}
\label{eq:beta_breathe}
\beta_{\rm intra}=\frac{\Delta_0^2}{2 s_0}\sum_{{\bf k}, a}\tau_{{\bf k}, a}\frac{\partial 
f_{{\bf k}, 
a}}{\partial\epsilon_{{\bf k}, a}}|\langle {\bf k}, a|K|{\bf k}, a\rangle|^2\frac{{\bf q}\cdot{{\bf v}_{d,a}}}{{\bf q}\cdot{\bf v}_s}
\end{equation}

\noindent where $K=[S^+,H_{so}]/\Delta_0$ is the spin-torque operator introduced in Eq. 
(~\ref{eq:kambersky}) and we have claimed spin rotational invariance via $|\langle[S^x,H_{so}]\rangle|^2=|\langle[S^y,H_{so}]\rangle|^2$. 
Using $\partial f/\partial\epsilon\simeq -\delta(\epsilon-\epsilon_F)$ and recalling from Section VII that $K_{a,a}=S^+_{a,a}$, Eq.~(\ref{eq:beta_breathe}) is equivalent to Eq.~(\ref{eq:intra_pre}); note that the product of spectral functions in the latter yields a factor of $4\pi\tau$ upon momentum integration.  
These observations prove that $\beta_{\rm intra}$ describes the contribution from a transport current to the ``breathing Fermi surface'' type of damping. Furthermore, Eq.~(\ref{eq:a_b_breathe}) highlights the importance of the ratio between the two characteristic velocities of a current carrying ferromagnet, namely $v_s$ and $v_d$. As explained in Section VI these two velocities coincide in models with Galilean invariance.  Only in these artificial models, which never apply to real materials, does $\alpha=\beta$ hold.

{\em Inter-band terms}: 
The Kohn-Sham orbitals are effective eigenstates of a mean-field Hamiltonian where the spins are aligned in the equilibrium direction.
As spins precess in response to external rf fields and dc transport currents, the time-dependent part of the mean-field Hamiltonian drives transitions between the ground-state Kohn-Sham orbitals. These processes lead to the second term in the effective field
and produce the inter-band contribution to damping.  

We thus concentrate on the second term in Eq.~(\ref{eq:h_eff_total}),
\begin{equation}
\label{eq:h_eff_inter}
{\cal H}^{\rm eff}_{{\rm inter}}=-\frac{1}{s_0}\sum_{{\bf k},a} \frac{\partial n_{{\bf k},a}}{\partial\hat\Omega}\epsilon_{{\bf k},a}.
\end{equation}
Multiplying Eq.~(\ref{eq:h_eff_inter}) with $\partial\hat\Omega/\partial t$ we get 

\begin{eqnarray}
\label{eq:h_eff_inter_b}
{\cal H}^{\rm eff, damp}_{\rm inter} \cdot \partial_t \hat{\Omega} &=&
-\frac{1}{s_0} \sum_{{\bf k},a} \epsilon_{{\bf k},a} \left [ \partial n_{a,{\bf k}} / \partial
\hat{\Omega} \cdot \partial \hat{\Omega} / \partial t \right ] \nonumber \\
 &=& -\frac{1}{s_0} \sum_{{\bf k},a} \epsilon_{{\bf k},a} \, \partial n_{a,{\bf k}} /
\partial t \, .
\end{eqnarray}

The rate of change of the populations of the Kohn-Sham states can be approximated by the following master equation

\begin{equation}
\label{eq:master2}
\frac{\partial n_{a,{\bf k}}}{\partial t} = -\sum_{b,{\bf k'}} W_{a,b} (n_{{\bf k},a}-n_{{\bf k'},b}),
\end{equation}
where 
\begin{equation}
\label{eq:rate}
W_{a,b} = 2\pi\left| \langle b,{\bf k'}|\Delta_0 S^x|a,{\bf k}\rangle\right |^2 \delta_{{\bf k'},{\bf k+q}} \delta(\epsilon_{b,{\bf k'}} - \epsilon_{a,{\bf k}} - \omega )
\end{equation}
is the spin-flip inter-band transition probability as dictated by Fermi's golden rule. Eqs.~ (\ref{eq:master2}) and ~(\ref{eq:rate}) rely on the principle of microscopic reversibility\cite{principle} and are rather {\em ad hoc} because they circumvent a rigorous analysis of the quasiparticle-magnon scattering, which would for instance require keeping track of magnon occupation number. Furthermore, quasiparticle-phonon and quasiparticle-impurity scattering are allowed for simply by broadening the Kohn-Sham eigenenergies (see below). 
The right hand side of Eq.~(\ref{eq:master2}) is now closely related to inter-band magnetization damping because it agrees\cite{pines} with the {\em net} decay rate of magnons into particle-hole excitations, where the particle and hole are in different bands. Combining Eq.~(\ref{eq:h_eff_inter_b}) and~(\ref{eq:master2}) and rearranging terms we arrive at
\begin{equation}
\label{eq:h_eff_inter2}
{\cal H}^{\rm eff}_{\rm inter}\cdot\partial_t \hat{\Omega} = \frac{1}{2 s_0}\sum_{{\bf k,k'},a,b} W_{a,b}(n_{{\bf k},a}-n_{{\bf k'},b})(\epsilon_{{\bf k},a}-\epsilon_{{\bf k'},b}).
\end{equation}
For the derivation of $\alpha_{\rm inter}$ it is sufficient to approximate $n_{{\bf k},a}$ as a Fermi distribution in Eq.~(\ref{eq:h_eff_inter2}); here we account for a transport current by shifting the Fermi seas as $n_{{\bf k},a}\to n_{{\bf k},a}-e {\bf v}_{{\bf k},a}\cdot{\bf E} \tau_{{\bf k},a} \partial n_{{\bf k},a}/\partial\epsilon_{{\bf k},a}$, which to leading order yields 
\begin{widetext}
\begin{eqnarray}
\label{eq:h_inter}
{\cal H}^{\rm eff}_{\rm inter} \cdot\partial_t \hat{\Omega}&=& -\frac{\pi\omega}{2 s_0} \sum_{{\bf k},a,b} \left| \langle b,{\bf k+q}|\Delta_0 S^+|a,{\bf k}\rangle\right |^2 
\delta(\epsilon_{b,{\bf k+q}} - \epsilon_{a,{\bf k}}-\omega)\frac{\partial n_{{\bf k},a}}{\partial \epsilon_{{\bf k},a}}\left(-\omega + e V_{b,a}\right)\nonumber\\
 &=& \frac{\omega}{8\pi s_0} \sum_{{\bf k},a,b} \left|\langle b,{\bf k+q}|\Delta_0 S^+|a,{\bf k}\rangle\right |^2 A_{a}({\bf k},\epsilon_F) A_{b}({\bf k+q},\epsilon_F) (-\omega+eV_{b,a})
\end{eqnarray}
where we have used $S^x=(S^+ +S^-)/2$ and defined $V_{b,a}=e {\bf v}_{{\bf k+q},b}\cdot{\bf E} \tau_{{\bf k+q},b}-e {\bf v}_{{\bf k},a}\cdot{\bf E} \tau_{{\bf k},a}$ . In the second line of Eq.(~\ref{eq:h_inter}) we have assumed low temperatures, and have introduced a finite quasiparticle lifetime by broadening the spectral functions of the Bloch states into Lorentzians with the convention outlined in Eq.~(\ref{eq:lorentzian}): $\delta(x)\to A(x)/(2\pi)$. 
Identifying Eq.(~\ref{eq:h_inter}) with  $(-\alpha_{\rm inter}\partial_t\hat\Omega-\beta_{\rm inter}({\bf v_s}\cdot\nabla)\hat\Omega)\cdot\partial_t\hat\Omega=-\alpha_{\rm inter}\omega^2+\beta_{\rm inter}\omega({\bf q}\cdot{\bf v_s})$ we arrive at 
\begin{eqnarray}
\alpha_{\rm inter} &=& \frac{1}{8\pi s_0} \sum_{a,b \neq a} \sum_{{\bf k},a,b} \left|\langle b,{\bf k+q}|\Delta_0 S^+|a,{\bf k}\rangle\right |^2 A_{a}({\bf k},\epsilon_F) A_{b}({\bf k+q},\epsilon_F)\nonumber\\
 \beta_{\rm inter} &=& \frac{1}{8\pi s_0 {\bf q}\cdot {\bf v_s}} \sum_{a,b \neq a} \sum_{{\bf k},a,b} \left|\langle b,{\bf k+q}|\Delta_0 S^+|a,{\bf k}\rangle\right |^2 A_{a}({\bf k},\epsilon_F) A_{b}({\bf k+q},\epsilon_F) V_{b,a}
\end{eqnarray}
in agreement with our results of Section II.
\end{widetext}

\section{Summary and Conclusions}
Starting from the Gilbert damping $\alpha$ and including the influence of an electric field in the transport orbitals semiclassically, we have proposed a concise formula for the non-adiabatic spin transfer torque coefficient $\beta$ that can be applied to real materials with arbitrary band structures.  Our formula for $\beta$ reproduces 
results obtained by more rigorous non-linear response theory calculations when applied to simple toy models. 
By applying this expression to a two-dimensional electron-gas ferromagnet with Rashba spin-orbit interaction, we have found that it 
implies a conductivity-like contribution to $\beta$, related to the corresponding contribution to the Gilbert damping $\alpha$,
which is proportional to scattering time rather than scattering rate and 
arises from intra-band transitions.  Our subsequent calculations using a four-band model have shown that intra-band contributions dominate in ferromagnetic semiconductors such as (Ga,Mn)As. We have then discussed the $\alpha/\beta$ ratio in realistic materials and have confirmed trends expected from toy models, in addition to suggesting that $\alpha$ and $\beta$ can have the opposite sign in systems where both hole-like and electron-like bands coexist at the Fermi surface. Afterwards, we have analyzed the spin-torque formalism suitable to \emph{ab-initio} calculations, and have concluded that it may provide a reliable estimate of the intra-band contribution to $\beta$; for the inter-band contribution the spin-torque formula offers a physically sensible result in the weak spin-orbit limit but its quantitative reliability is questionable unless the prominent inter-band transitions connect states that are close in energy. Finally, we have extended the breathing Fermi surface model for the Gilbert damping to current carrying ferromagnets and have accordingly found a complementary physical interpretation for the intra-band contribution to $\beta$; similarly, we have applied the master equation in order to offer an alternative interpretation for the inter-band contribution to $\beta$. Possible avenues for future research consist of carefully analyzing the importance of higher order vertex corrections in $\beta$, better understanding the disparities between the different approaches to $v_s$, and finding real materials where $\alpha/\beta$ is negative. 
 
\section*{Acknowledgements}
We acknowledge informative correspondence with Rembert Duine and
Hiroshi Kohno. In addition, I.G. is grateful to Paul Haney for
interesting discussions and generous hospitality during his stay in
the National Institute of Standards and Technology. 
This work was supported in part by the Welch Foundation, by the National
Science Foundation under grant DMR-0606489, and by the
NIST-CNST/UMD-NanoCenter Cooperative Agreement.

\appendix
\begin{widetext}
\section{Quadratic Spin Response to an Electric and Magnetic Field}
Consider a system that is perturbed from equilibrium by a time-dependent perturbation ${\cal V}(t)$. 
The change in the expectation value of an operator $O(t)$ under the influence of  ${\cal V}(t)$ can be formally expressed as 
\begin{equation}
\delta \langle O(t)\rangle=\langle \Psi_0|U^{\dagger}(t)O(t)U(t)|\Psi_0\rangle-\langle\Psi_0|O(t)|\Psi_0\rangle
\end{equation}
where $|\Psi_0\rangle$ is the unperturbed state of the system,
\begin{equation}
U(t)=T \exp\left[-i\int_{-\infty}^t {\cal V}(t^{\prime}) d t^{\prime}\right]
\end{equation}
is the time-evolution operator in the interaction representation and $T$ stands for time ordering. Expanding the exponentials up to second order in ${\cal V}$ we arrive at
\begin{equation}
\label{eq:quadratic response}
\delta \langle O(t)\rangle=i\int_{-\infty}^t dt'\langle\left[O(t),{\cal V}(t')\right]\rangle-\frac{1}{2}\int_{-\infty}^t dt' dt''\langle\left[\left[O(t),{\cal V}(t')\right],{\cal V}(t'')\right]\rangle.
\end{equation}
\begin{figure}[b]
\begin{center}
\scalebox{0.4}{\includegraphics{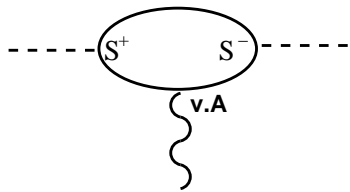}}
\caption{Feynman diagram for $\chi_{S,S,j}$. The dashed lines correspond to magnons, whereas the wavy line represents a photon.}
\label{fig:quadratic}
\end{center}
\end{figure}
For the present work, $O(t)\to S^a$ ($a=x,y,z$) and 
\begin{equation}
\label{eq:V}
{\cal V}(t)=-\int d{\bf r} {\bf j}\cdot {\bf A}({\bf r},t)+\int d{\bf r} {\bf S}\cdot {\bf {\cal H}_{\rm ext}}({\bf r},t),
\end{equation}
where ${\bf A}$ is the vector potential, ${\bf {\cal H}_{\rm ext}}$ is the external magnetic field, and ${\bf j}$ is the current operator. Plugging Eq.~(\ref{eq:V}) into Eq.~(\ref{eq:quadratic response}) and neglecting $O(A^2), O({\cal H}_{\rm ext}^2)$ terms we obtain
\begin{equation}
\delta S^a (x)=\sum_b\int dx' \chi^{a,b}_{S,j} A^b (x')+\sum_b\int dx' \chi^{a,b}_{S,S} {\cal H}_{\rm ext}^b (x')+\sum_{b,c}\int dx' dx'' \chi^{a,b,c}_{S,S,j} A^b (x') {\cal H}_{\rm ext}^c (x''),
\end{equation}
where $x\equiv ({\bf r},t)$ and $\int dx'\equiv \int_{-\infty}^{\infty} dt' \int d{\bf r}'$. The linear and quadratic response functions introduced above are defined as
\begin{eqnarray}
\chi^{a,b}_{S,j}(x,x')&=& i\langle\left[S^a(x),j^b(x')\right]\Theta(t-t')\nonumber\\
\chi^{a,b}_{S,S}(x,x')&=& i\langle\left[S^a(x),S^b(x')\right]\Theta(t-t')\nonumber\\
\chi^{a,b,c}_{S,S,j}(x,x',x'')&=& \langle\left[\left[S^a(x),j^b(x')\right],S^{c}(x'')\right]\Theta(t-t')\Theta(t'-t'')\nonumber\\
&&+ \langle\left[\left[S^a(x),S^b(x'')\right],j^{c}(x')\right]\Theta(t-t'')\Theta(t''-t')
\end{eqnarray}
where we have used $T\left[F(t) G(t')\right]=F(t') G(t'') \Theta(t'-t'')+G(t'') F(t')\Theta(t''-t')$, $\Theta$ being the step function. $\chi_{S,j}$ is the spin density induced by an electric field in a uniform ferromagnet, and it vanishes unless there is intrinsic spin-orbit interaction. $\chi_{S,S}$ is the spin density induced by an external magnetic field. $\chi_{S,S,j}$ is the spin density induced by the combined action of an electric and magnetic field (see Fig.~(\ref{fig:quadratic}) for a diagrammatic representation); this quantity is closely related to $({\bf v}_{\rm s}\cdot {\bf q} )\chi^{(2)}$, introduced in Section II.

\section{First order impurity vertex correction}
The aim of this Appendix is to describe the derivation of  Eq.~(\ref{eq:beta_vertex}). We shall begin by evaluating the leading order vertex correction to the Gilbert damping. From there, we shall obtain the counterpart quantity for the non-adiabatic STT by shifting the Fermi occupation factors to first order in the electric field.
\begin{figure}
\begin{center}
\scalebox{0.4}{\includegraphics{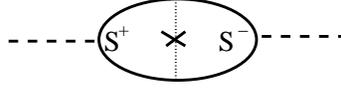}}
\caption{Feynman diagram for the first order vertex correction. The dotted line with a cross represents the particle-hole correlation mediated by impurity scattering. }
\label{fig:first vertex}
\end{center}
\end{figure}
The analytical expression for the transverse spin response with one vertex correction is (see Fig.~(\ref{fig:first vertex}))
\begin{equation}
\tilde{\chi}_{+,-}^{{\rm QP},(1)}=- V\frac{\Delta_0^2}{2} T\sum_{\omega_{n}}\int_{\textbf{k},\textbf{k}^{\prime}} u^{i} G_{a}(i\omega_{n},\textbf{k})S^{+}_{a,b}G_{b}(i\omega_{n}+i\omega,\textbf{k}+\textbf{q})S^{i}_{a,b^{\prime}} G_{b^{\prime}}(i\omega_{n}+i\omega,\textbf{k}^{\prime}+\textbf{q})S^{-}_{b^{\prime},a^{\prime}} G_{a^{\prime}}(i\omega_{n},\textbf{k}^{\prime})S^{i}_{a^{\prime},a}.
\end{equation}
where $V$ is the volume of the system and the minus sign originates from fermionic statistics. Using the Lehmannn representation of the Green's functions $G$ and performing the Matsubara sum we get
\begin{eqnarray}
\tilde{\chi}_{+,-}^{{\rm QP},(1)}&=&-V\frac{\Delta_0^2}{2}\int_{\textbf{k},\textbf{k}^{\prime}} u^{i} 2\mbox{ Re }\left[S^{+}_{a,b}S^{i}_{b,b^{\prime}}S^{-}_{b^{\prime},a^{\prime}}S^{i}_{a^{\prime},a}\right]  \int_{-\infty}^{\infty}\frac{d\epsilon_{1} d\epsilon_{1}^{\prime} d\epsilon_{2} d\epsilon_{2}^{\prime}}{(2\pi)^{4}} A_{a}(\epsilon_{1},\textbf{k}) A_{a^{\prime}}(\epsilon_{1}^{\prime},\textbf{k}^{\prime}) \nonumber\\
&&\times 
A_{b}(\epsilon_{2},\textbf{k}+\textbf{q})
A_{b^{\prime}}(\epsilon_{2}^{\prime},\textbf{k}^{\prime}+\textbf{q})
\left[\frac{f(\epsilon_{1})}{(\epsilon_{1}-\epsilon_{1}^{\prime})(i\omega+\epsilon_{1}-\epsilon_{2})(i\omega+\epsilon_{1}-\epsilon_{2}^{\prime})}+\left(\begin{array}{c}\epsilon_{1}\leftrightarrow\epsilon_{2},\epsilon_{1}^{\prime}\leftrightarrow\epsilon_{2}^{\prime},\\\omega\leftrightarrow -\omega\end{array}\right)\right]
\end{eqnarray}
where twice the real part arose after absorbing two of the terms coming from the Matsubara sum. Next, we apply $i\omega\rightarrow\omega+i 0^{+}$ and take the imaginary part: 
\begin{eqnarray}
\tilde{\chi}_{+,-}^{{\rm QP},(1)}&=&V\frac{\Delta_0^2}{2} 2\pi\int_{\textbf{k},\textbf{k}^{\prime}} u^{i} \mbox{Re }\left[S^{+}_{a,b}S^{i}_{b,b^{\prime}}S^{-}_{b^{\prime},a^{\prime}}S^{i}_{a^{\prime},a}\right]  \int_{-\infty}^{\infty}\frac{d\epsilon_{1} d\epsilon_{1}^{\prime} d\epsilon_{2} d\epsilon_{2}^{\prime}}{(2\pi)^{4}} A_{a}(\epsilon_{1},\textbf{k}) A_{a^{\prime}}(\epsilon_{1}^{\prime},\textbf{k}^{\prime}) A_{b}(\epsilon_{2},\textbf{k}+\textbf{q}) A_{b^{\prime}}(\epsilon_{2}^{\prime},\textbf{k}^{\prime}+\textbf{q})\nonumber\\
&&\times \frac{f(\epsilon_{1})}{\epsilon_{1}-\epsilon_{1}^{\prime}}
\left[\frac{\delta(\omega+\epsilon_{1}-\epsilon_{2})}{\omega+\epsilon_{1}-\epsilon_{2}^{\prime}}+\frac{\delta(\omega+\epsilon_{1}-\epsilon_{2}^{\prime})}{\omega+\epsilon_{1}-\epsilon_{2}}-\left(\begin{array}{c}\omega\rightarrow -\omega,\\ \textbf{q}\rightarrow -\textbf{q}\end{array}\right)\right]
\end{eqnarray}
where we used $1/(x-i\eta)=PV(1/x)+i\pi\delta(x)$, and invoked spin-rotational invariance to claim that terms with $S^{x}_{a,b}S^{i}_{b,b^{\prime}}S^{y}_{b^{\prime},a^{\prime}}S^{i}_{a^{\prime},a}$ will vanish. Integrating the delta functions we arrive at
\begin{eqnarray}
\tilde{\chi}_{+,-}^{{\rm QP},(1)}&=&V\frac{\Delta_0^2}{2}\int_{\textbf{k},\textbf{k}^{\prime}}u^{i}\mbox{Re}\left[...\right]\int_{-\infty}^{\infty}\frac{d\epsilon_{1}^{\prime} d\epsilon_{2} d\epsilon_{2}^{\prime}}{(2\pi)^{3}}\frac{f(\epsilon_{2})A_{a}(\epsilon_{2},\textbf{k})A_{a^{\prime}}(\epsilon_{1}^{\prime},\textbf{k}^{\prime})}{(\epsilon_{2}-\epsilon_{2}^{\prime})(\epsilon_{2}-\epsilon_{1}^{\prime})}\nonumber\\
&&\times\left[A_{b}(\epsilon_{2}+\omega,\textbf{k}+\textbf{q}) A_{b^{\prime}}(\epsilon_{2}^{\prime}+\omega,\textbf{k}^{\prime}+\textbf{q})+A_{b}(\epsilon_{2}^{\prime}+\omega,\textbf{k}+\textbf{q}) A_{b^{\prime}}(\epsilon_{2}+\omega,\textbf{k}^{\prime}+\textbf{q})\right] -\left(\begin{array}{c}\omega\rightarrow -\omega,\\ \textbf{q}\rightarrow -\textbf{q}\end{array}\right)
\end{eqnarray}

The next step is to do the $\epsilon_{1}^{\prime}$ and $\epsilon_{2}^{\prime}$ integrals, taking advantage of the fact that for weak disorder the spectral functions are sharply peaked Lorentzians ( in fact at the present order of approximation one can take regard them as Dirac delta functions). The result reads 
\begin{eqnarray}
\tilde{\chi}_{+,-}^{{\rm QP},(1)}&=&V\frac{\Delta_0^2}{2}\int_{\textbf{k},\textbf{k}^{\prime}}u^{i}\mbox{Re}\left[...\right]\int_{-\infty}^{\infty}\frac{d\epsilon_{2}}{2\pi}
\frac{f(\epsilon_{2})A_{a}(\epsilon_{2},\textbf{k})}{\epsilon_{2}-\epsilon_{\textbf{k}^{\prime},a^{\prime}}}\left[\frac{A_{b}(\epsilon_{2}+\omega,\textbf{k}+\textbf{q})}{\epsilon_{2}+\omega-\epsilon_{\textbf{k}^{\prime}+\textbf{q},b^{\prime}}}+\frac{A_{b^{\prime}}(\epsilon_{2}+\omega,\textbf{k}^{\prime}+\textbf{q})}{\epsilon_{2}+\omega-\epsilon_{\textbf{k}+\textbf{q},b}}\right] \nonumber\\
&&-\left(\omega\rightarrow -\omega, \textbf{q}\rightarrow -\textbf{q}\right)
\end{eqnarray}
By making further changes of variables, this equation can be rewritten as
\begin{equation}
\label{eq:alpha_vertex}
\tilde{\chi}_{+,-}^{{\rm QP},(1)}=V\frac{\Delta_0^2}{2}\int_{\textbf{k},\textbf{k}^{\prime}}u^{i}\mbox{Re}\left[...\right]\int_{-\infty}^{\infty}\frac{d\epsilon_{2}}{2\pi}
\frac{\left(f(\epsilon_{2})-f(\epsilon_{2}+\omega)\right)A_{a}(\epsilon_{2},\textbf{k})}{\epsilon_{2}-\epsilon_{\textbf{k}^{\prime},a^{\prime}}}\left[\frac{A_{b}(\epsilon_{2}+\omega,\textbf{k}+\textbf{q})}{\epsilon_{2}+\omega-\epsilon_{\textbf{k}^{\prime}+\textbf{q},b^{\prime}}}+\frac{A_{b^{\prime}}(\epsilon_{2}+\omega,\textbf{k}^{\prime}+\textbf{q})}{\epsilon_{2}+\omega-\epsilon_{\textbf{k}+\textbf{q},b}}\right]
\end{equation}
This is the first order vertex correction for the Gilbert damping. In order to obtain an analogous correction for the non-adiabatic STT, it suffices to shift the Fermi factors in Eq.~(\ref{eq:alpha_vertex}) as indicated in the main text. This immediately results in Eq.~(\ref{eq:beta_vertex}). 

\section{Derivation of Eq.~(\ref{eq:beta_M2DEG})}

Let us first focus on the first term of Eq.~(\ref{eq:beta_linearized}), namely
\begin{equation}
E_{i} q_{j}\int_{\textbf{k}}\left[|\langle a,\textbf{k}|S^{+}|b,\textbf{k}\rangle|^{2}+|\langle a,\textbf{k}|S^{-}|b,\textbf{k}\rangle|^{2}\right] A_{a} A_{b}^{\prime} v^{i}_{\textbf{k},a} v^{j}_{\textbf{k},b}\tau_{\textbf{k},a}
\end{equation}
We shall start with the azimuthal integral.  It is easy to show that the entire angle dependence comes from $v^{i}v^{j}\propto k_{i} k_{j}$, from which the azimuthal integral vanishes unless $i=j$.\\
Regarding the $|k|$ integral, we assume that $\lambda k_{\rm F}, \Delta_0, 1/\tau << \epsilon_{\rm F}$; otherwise the analytical calculation is complicated and must be tackled numerically. Such assumption allows us to use $\int_{\textbf{k}}\rightarrow N_{2D}\int_{-\infty}^{\infty} d\epsilon$. For inter-band transitions ($a\neq b$), $A_{a}A^{\prime}_{b}$ contributes mainly thru the pole at $\epsilon_{{\rm F},a}$, thus all the slowly varying factors in the integrand may be set at the Fermi energy. For intra-band transitions ($a=b$), $A_{a}A^{\prime}_{a}$ has no peak at the Fermi energy; hence it is best to keep the slowly varying factors inside the integrand.

The above observations lead straightforwardly to the following result:
\begin{eqnarray}
\label{eq:M2DEG_integral_1}
&&E_{i} q_{j}\int_{\textbf{k}}\left[|\langle a,\textbf{k}|S^{+}|b,\textbf{k}\rangle|^{2}+|\langle a,\textbf{k}|S^{-}|b,\textbf{k}\rangle|^{2}\right] A_{a} A_{b}^{\prime} v^{i}_{\textbf{k},a} v^{j}_{\textbf{k},b}\tau_{\textbf{k},a}\nonumber\\
&\simeq&\textbf{E}\cdot\textbf{q}\frac{m^{2}}{8 m_{+} m_{-}}\left(1+\frac{\Delta_0^{2}}{b^{2}}\right)\frac{\left(\epsilon_{{\rm F},-}\tau_{-}\Gamma_{+}-\epsilon_{{\rm F},+}\tau_{+}\Gamma_{-}\right)}{b^{3}}\nonumber\\
&-&\textbf{E}\cdot\textbf{q}\left[\frac{m^{2}}{m_{+}^{2}}\frac{1}{2}\frac{\lambda^{2}k_{\rm F}^{2}}{b^{2}}\left(1+\frac{\Delta_0^{2}}{b^{2}}\right)\tau_{+}^{2}+\frac{m^{2}}{m_{-}^{2}}\frac{1}{2}\frac{\lambda^{2}k_{\rm F}^{2}}{b^{2}}\left(1+\frac{\Delta_0^{2}}{b^{2}}\right)\tau_{-}^{2}\right]
\end{eqnarray}
The second and third line in Eq.~(\ref{eq:M2DEG_integral_1}) come from inter-band  and intra-band transitions, respectively. The latter vanishes in absence of spin-orbit interaction, leading to a 2D version of Eq.~(\ref{eq:stoner_2}). Since the band-splitting is much smaller than the Fermi energy, one can further simplify the above equation via $\tau_{+}\simeq\tau_{-}\rightarrow \tau$.

Let us now move on the second term of Eq.~(\ref{eq:beta_linearized}), namely
\begin{equation}
E_{i}q_{j}\int_{\textbf{k}}\mbox{Re}\left[\langle b,\textbf{k}|S^{-}|a, \textbf{k}\rangle\langle a,\textbf{k}|S^{+}\partial_{k_{j}}|b,\textbf{k}\rangle +(S^{+}\leftrightarrow S^{-})\right]A_{a}A_{b}v^{i}_{\textbf{k},a}\tau_{\textbf{k},a}
\end{equation} 
Most of the observations made above apply for this case as well. For instance, the azimuthal integral vanishes unless $i=j$. This follows from a careful evaluation of the derivatives of the eigenstates with respect to momentum; $\partial_{k_{j}}\theta=\sin(\theta)\cos(\theta) k_{j}/k^{2}$ ($0\leq\theta\leq\pi/2$) is a useful relation in this regards, while $\partial_{k_{j}}\phi$ plays no role. As for the $|k|$ integral, we no longer have the derivative of a spectral function, but rather a product of two spectral functions; the resulting integrals may be easily evaluated using the method of residues. The final result reads
\begin{eqnarray}
\label{eq:M2DEG_integral_2}
&&E_{i}q_{j}\int_{\textbf{k}}\mbox{Re}\left[\langle b,\textbf{k}|S^{-}|a, \textbf{k}\rangle\langle a,\textbf{k}|S^{+}\partial_{k_{j}}|b,\textbf{k}\rangle +(S^{+}\leftrightarrow S^{-})\right]A_{a}A_{b}v^{i}_{\textbf{k},a}\tau_{\textbf{k},a}\nonumber\\
&\simeq&-\textbf{E}\cdot\textbf{q}\left[\frac{m}{32 m_{-}}\frac{\lambda^{2}k_{\rm F}^{2}\Delta_0^{2}}{b^{6}}\left(1+\frac{\tau_{-}}{\tau_{+}}\right)+\frac{m}{32 m_{+}}\frac{\lambda^{2}k_{\rm F}^{2}\Delta_0^{2}}{b^{6}}\left(1+\frac{\tau_{+}}{\tau_{-}}\right)\right]\nonumber\\
&+&\textbf{E}\cdot\textbf{q}\left[\frac{m}{4 m_{+}}\frac{\lambda^{2}k_{\rm F}^{2}\Delta_0^{2}}{b^{4}}\tau_{+}^{2}+\frac{m}{4 m_{-}}\frac{\lambda^{2}k_{\rm F}^{2}\Delta_0^{2}}{b^{4}}\tau_{-}^{2}\right]
\end{eqnarray}
The first line in Eq.~(\ref{eq:M2DEG_integral_2}) stems from inter-band transitions, whereas the second comes from intra-band transitions; \emph{both} vanish in absence of SO. Once again we can take  $\tau_{+}\simeq\tau_{-}\rightarrow \tau$. Combining Eqs.~(\ref{eq:M2DEG_integral_1}) and~(\ref{eq:M2DEG_integral_2}) one can immediately reach Eq.~(\ref{eq:beta_M2DEG}).
\end{widetext}


\begin{thebibliography}{30}
\bibitem{convention} Here we assume that the dependence of energy on magnetization direction which determines 
${\cal H}_{\rm eff}$ is specified as a function of $\Omega_x$ and $\Omega_y$ only with 
$\Omega_z$ implicitly fixed by the constraint $\Omega_z = [1 - \Omega_x^2 - \Omega_y^2]^{1/2}$. 
If the free energy was expressed in a form with explicit $\Omega_z$ dependence we would find  
${\cal H}_{{\rm eff},x}=-\partial{F}/\partial{\Omega_x}-(\partial{F}/\partial{\Omega_z})(\partial\Omega_z/\partial\Omega_x)=-\partial{F}/\partial{\Omega_x}+(\partial{F}/\partial{\Omega_z})\Omega_x$, where $F$ is the free energy of the ferromagnet. Similarly we would find 
${\cal H}_{{\rm eff},y}=-\partial F/\partial\Omega_y+(\partial F/\partial\Omega_z)\Omega_y$.  The terms which arise from the $\Omega_z$ dependence of the free energy would more commonly be regarded as contributions to ${\cal H}_{eff,z}$.  The difference is purely a matter 
of convention since both results would give the same value for $\hat{\Omega} \times {\cal H}_{\rm eff}$.
\bibitem{xx_vs_pm} For convenience in Eq.~(\ref{eq:alpha_beta_SDFT}) we use $\langle S^+ S^-\rangle$ response functions instead of $\langle S^x S^x\rangle$ and $\langle S^y S^y\rangle$. They are related via $S^x=(S^+ + S^-)/2$ and $S^y=(S^+ - S^-)/2i$.
\bibitem{exact} In order to gauge the accuracy of either matrix element, one must obtain an exact evaluation of the non-adiabatic STT, which entails a ladder-sum renormalization\cite{alphaII} of $S^{\pm}$. This is beyond the scope of the present work.
\bibitem{principle} This principle states that $W_{a,b}=W_{b,a} \exp((\epsilon_a-\epsilon_b)/T)$. Since the magnon energy is much smaller than the uncertainty in the quasiparticle energies, we approximate $W_{a,b}\simeq W_{b,a}$.
\end{thebibliography}
\end{document}